\documentclass{aastex61}

\begin{document}
\title{The formation and decay of sunspot penumbra in Active Region NOAA 12673}

\email{yanxl@ynao.ac.cn}

\author{Qiaoling Li}
\affiliation{Yunnan Observatories, Chinese Academy of Sciences, Kunming 650011, China}
\affiliation{University of Chinese Academy of Sciences, Yuquan Road, Shijingshan Block Beijing 100049, China.}

\author{Xiaoli Yan}
\affiliation{Yunnan Observatories, Chinese Academy of Sciences, Kunming 650011, China}
\affiliation{Center for Astronomical Mega-Science, Chinese Academy of Sciences, 20A Datun Road, Chaoyang District, Beijing, 100012, China.}

\author{Jincheng Wang}
\affiliation{Yunnan Observatories, Chinese Academy of Sciences, Kunming 650011, China}
\affiliation{Center for Astronomical Mega-Science, Chinese Academy of Sciences, 20A Datun Road, Chaoyang District, Beijing, 100012, China.}

\author{DeFang Kong}
\affiliation{Yunnan Observatories, Chinese Academy of Sciences, Kunming 650011, China}
\affiliation{Center for Astronomical Mega-Science, Chinese Academy of Sciences, 20A Datun Road, Chaoyang District, Beijing, 100012, China.}

\author{Zhike Xue}
\affiliation{Yunnan Observatories, Chinese Academy of Sciences, Kunming 650011, China}
\affiliation{Center for Astronomical Mega-Science, Chinese Academy of Sciences, 20A Datun Road, Chaoyang District, Beijing, 100012, China.}

\author{Liheng Yang}
\affiliation{Yunnan Observatories, Chinese Academy of Sciences, Kunming 650011, China}
\affiliation{Center for Astronomical Mega-Science, Chinese Academy of Sciences, 20A Datun Road, Chaoyang District, Beijing, 100012, China.}
\begin{abstract}
To better understand the formation and decay of sunspot penumbra, we studied the evolution of sunspots in the three regions of the active-region NOAA 12673 in detail. The evolution of sunspots in the three regions were all involved in the interaction of two magnetic field systems: the pre-existing magnetic field system and the later emerging magnetic field system. Through analyzing the photospheric magnetic field properties, it is found that the formation of the penumbra originated from the newly emerging magnetic bipole that were trapped in the photosphere. The change of magnetic fields in penumbra from horizontal to vertical can cause the disappearance of penumbra. The transformation of the magnetic field between the umbra and the penumbra is found and the outward moat flow around sunspot gradually decreased and vanished during the decay of sunspot. In addition, we found that the mean longitudinal magnetic strength in penumbra decreased and the mean transverse magnetic strength in penumbra increased with the increasing penumbral area during the formation of sunspots. However, during the decay of sunspots, the mean longitudinal magnetic strength in penumbra increased and the mean transverse magnetic strength in penumbra decreased with the decreasing penumbral area. Comparatively, the dependence of the area and the mean transverse/longitudinal magnetic field strength in umbra is not remarkable. These results reveal that the formation and decay process of umbra are different with penumbra.
\end{abstract}

\section{Introduction}\label{sec:introduction}
Sunspots are the most noticeable manifestation of solar magnetic field concentrations in the photosphere. A typical mature sunspot has umbra and penumbra. The presence of a penumbra distinguishes a mature sunspot from the small pore. The formation and decay of sunspot penumbra is a considerably complex process. Although the formation and decay of sunspot penumbrae occur frequently in the active region, the formation mechanism and the origination of the penumbral magnetic field are poorly understood  due to the lack of observations that fully cover the penumbral evolution process with high spatial, spectral and temporal resolution.

When sunspot penumbra begins to form, the transitions in the area and the magnetic field from umbra to penumbra seem to have a impact on the development of penumbral filaments. As the total magnetic flux of a pore increases, the outermost magnetic field lines of pore become more horizontal because of the demand of the force balance. Once these field lines incline to a critical angle, a convectively driven filamentary instability sets in \citep{Hurlburt..2000MNRAS.314..793H,Tildesley..2004MNRAS.350..657T}. Then the field lines at the boundary of pore are so horizontal that they are grabbed and pumped downwards by the surrounding granular convection. The development of this process leads to the formation of the penumbra with its interlocking-comb field. This mechanism for penumbral formation is referred to as ``flux pumping'' \citep{Thomas..2002Natur.420..390T,Weiss..2004ApJ...600.1073W,Brummell..2008ApJ...686.1454B}. Using Fe \uppercase\expandafter{\romannumeral1} 630.25 nm spectral line, \cite{Romano..2014ApJ...784...10R} detected some patches around a pore, in which showed upward motions and exhibited radially outward displacement. These features were interpreted as the footpoints of the pore magnetic field lines. They suggested that the magnetic field lines rooted in the pore were forced to return to the photosphere by the chromospheric magnetic fields and were progressively stretched and pushed down by the granular convection. The penumbra forms as a result of the change in inclination of magnetic field lines of umbra (from a vertical to a horizontal configuration), as supported by the investigation of \cite{Murabito..2016ApJ...825...75M}.

Based on the study of ten sunspots, \cite{Jur..2011A&A...531A.118J} analysed the umbra-penumbra boundary and concluded that the inner stable penumbral boundary are defined by the critical value of the vertical component of the sunspot magnetic field ($B^{ver}_{stable}$ = 1.8 KG). If the vertical component of magnetic field ($B_{ver}$) in pore was smaller than $B^{ver}_{stable}$, the penumbral magneto-convective mode in the pore would be not hindered and a stable pore-penumbra boundary would not be established. \cite{Jur..2017A&A...597A..60J} further analyzed the penumbral formation of a pore and confirmed the necessity of $B^{ver}_{stable}$ for establishing a stable umbra-penumbra boundary \citep{Jur..2015A&A...580L...1J}. They found that the penumbra grew at the expense of magnetic flux of the pore, which supports the result proposed earlier by \cite{Watanabe..2014ApJ...796...77W}. When the rudimentary penumbral filaments formed, \cite{Watanabe..2014ApJ...796...77W} observed that the area of the dark umbra gradually decreased. However, during the process of penumbral decay, they found that the horizontal penumbral field became vertical resulting in the recovery of the umbral area. Conversely, \cite{Schlichenmaier 2010AN....331..563S} found that the umbral area remained a constant value while a sunspot penumbra formed.

According to some studies on penumbral formation, the magnetic field in the nearby sunspots, such as flux emergence and the pre-existing overlying magnetic field, plays a pivotal role in the formation of penumbra. The emergence of magnetic flux in the vicinity of a sunspot can lead to a critical point in penumbral formation\citep{Zwaan..1992ASIC..375...75Z,Leka..1998ApJ...507..454L,Yang..2003ApJ...597.1190Y,Zuccarello..2014ApJ...787...57Z}. \cite{Leka..1998ApJ...507..454L} suggested that the emerging horizontal field lines would be trapped in the photosphere and form penumbral filaments rather than continuing to rise to higher layers. However, how the emerging field is trapped in the photosphere is still not clear. \cite{Shimizu..2012ApJ...747L..18S} suggested that the chromospheric canopy magnetic structure may play an important role in the process of emerging horizontal fields trapped in the photosphere. They noticed the presence of an annular zone of 3$^\prime$$^\prime$ -- 5$^\prime$$^\prime$ width in Ca ${\uppercase\expandafter{\romannumeral2}}H$ images around pore before the formation of its penumbra. The annular zone reflects the formation of a magnetic field overlying the surrounding of pore at the chromosphere. \cite{Lim..2013ApJ...769L..18L} observed some chromospheric threads above the forming penumbral filaments in an emerging flux region. They thought the chromospheric threads represent the pre-existing chromospheric horizontal field during the formation of penumbra filaments and suggested the emerging flux is trapped at the photosphere by the overlying chromospheric canopy fields. The presence of magnetic canopy fields in the chromosphere before penumbra formation may be crucial for the development of penumbra evidenced by chromospheric observations \citep{Romano..2013ApJ...771L...3R,Romano..2014ApJ...784...10R,Guglielmino..2014ApJ...786L..22G} and simulations \citep{Rempel..2011ApJ...729....5R,Rempel..2012ApJ...750...62R,MacTaggart..2016ApJ...831L...4M}. However, not all events of penumbral formation is accompanied with the overlying magnetic field before their formation. The effect of overlying magnetic field on penumbral filaments deserved further study.

Despite the emerging magnetic flux may supply the additional magnetic flux to penumbra, some investigations have an opposite point of view. The emerging flux seems to prevent the formation of steady penumbra. \cite{Schlichenmaier..2010A&A...512L...1S} and \cite{Rezaei..2012A&A...537A..19R} studied a same sunspot in the active region NOAA 11024, and noted that the stable penumbral sector formed on the side away from the flux emergence region. Both studies agreed that the ongoing flux emergence around sunspot inhibited the formation of stable penumbra and suggested a ``quiet'' magnetic surrounding in the vicinity of a sunspot seems to facilitate the formation of stable penumbra on a dynamical timescale. In contrast, \cite{Louis..2013A&A...552L...7L} found some stable penumbral filaments of a decaying sunspot formed after the newly emerged patch coalesced with the sunspot. Likewise, \cite{Murabito..2017ApJ...834...76M} observed that the first stable penumbral sector around a pore formed in the flux emergence region. By analyzing the penumbral formation in the twelve active regions, \cite{Murabito..2018ApJ...855...58M} reported that eight sunspots formed the first stable penumbral sector in the flux emergence and nine sunspots formed on the side away from the flux emergence. Consequently, these contradictory results about penumbral formation in flux emergence lead to an open question: what role does emerging flux play in the penumbral formation?

Penumbral decay is a relatively slow process compared with the process of penumbral formation. In particular, there are many events in which the penumbra quickly disappears. These events mostly are related to solar flares or other eruptive phenomena. Some studies found the penumbra rapidly decay after some X-class solar flares because the magnetic fields of the penumbra turned from horizontal to vertical \citep{Wang..2004ApJ...601L.195W,Deng..2005ApJ...623.1195D}. \cite{Bellot Rubio..2008ApJ...676..698B} observed a sunspot that gradually lost its penumbra in three days and discovered some finger-like structures near the decaying sunspot. These features were characterized by the weak horizontal magnetic field and blueshift. The authors speculated these structures may be related to penumbral magnetic field lines, which rise to the chromosphere by buoyancy and result in the disappearance of penumbra in the photosphere. By using the high-resolution observations of a decaying sunspot, \cite{Verma..2018A&A...614A...2V} found that the horizontal magnetic fields in penumbra became vertical when the sunspot decayed. This conclusion is same as the explanation of flare-induced rapid penumbral decay.

Except for the changes of magnetic field of penumbra, the magnetic flux removal of sunspot is an explanation of sunspot decay. Moving magnetic features (MMFs) are a key feature of the magnetic flux dispersal process and are only associated with decaying sunspots \citep{Harvey..1973SoPh...28...61H}. \cite{Mart Pillet..2002AN....323..342M} analyzed the decay of leading sunspots and trailing sunspots and explained how magnetic flux of sunspot was spread over a larger area when the sunspot was decaying, but failed to satisfactorily describe the origin of flux removal process. It has been shown that the sunspot flux removal process mainly includes fragmentation of the umbra, flux cancelation of MMFs and flux transport by MMFs to the surrounding region, as suggested by \cite{Deng..2007ApJ...671.1013D}. Similarly, \cite{Verma..2012A&A...538A.109V} found that magnetic flux was carried by MMFs from the decaying sunspot. However, the relation between MMFs and the decay of penumbra is still a matter of debate \citep{Cabrera..2006ApJ...649L..41C}.

\cite{Rempel..2015ApJ...814..125R}  have analyzed how sunspot decay by numerical simulation. In his simulations, he found two factors that inhibited sunspot decay. One is a strong reduction of the downflow filling factor and convective rms velocity underneath the sunspot penumbra. The other is the outer boundary of the naked pore. The reduction of the downflow filling factor prevents the submergence of horizontal magnetic field, which turns out to be the dominant decay process in the simulations. In addition, they noted that the deeper seated convective motions perhaps can erode the ``footpoint" of the sunspot and lead to flux separation resulting in the decay of the sunspot.

In this paper, we study the evolution of sunspots in three regions of active region NOAA 12673. We focus on how the emerging flux affect penumbral formation and decay. The paper is organized as follows: the observations are described in Section \ref{sec:observations}. The details of the results are presented in Section \ref{sec:results}. The conclusion are given in Section \ref{sec:conclusion} and in Section \ref{sec:discussion} we discuss some of the findings.

\section{Observations}\label{sec:observations}
The main data used in this paper are the full-disk continuum intensity images and line-of sight (LOS) magnetograms with a 45 s cadence and a pixel scale of 0.5$^\prime$$^\prime$ taken by the Helioseismic and Magnetic Imager (HMI; \cite{Schou..2012SoPh..275..229S}) on board the Solar Dynamic Observatory (SDO; \cite{Scherrer..2012SoPh..275..207S}). The continuum intensity images and LOS magnetograms are used to show the temporal evolution of sunspots in active region NOAA 12673 from 2017 September 2 to September 6.

To study the evolution of the magnetic field, we also analyzed the Space-weather HMI Active Region Patches (SHARP) vector magnetogram data \citep{Bobra..2014SoPh..289.3549B,Centeno..2014SoPh..289.3531C}.  The SHARP data are generated from the polarization measurements at six wavelengths along the Fe \uppercase\expandafter{\romannumeral1} 617.3 nm  spectral line \citep{Hoeksema..2014SoPh..289.3483H} and are inverted by using the Very Fast Inversion of the Stokes Vector (VFISV) algorithm \citep{Borrero..2011SoPh..273..267B} based on the Milne-Eddington approximation. The $180\,^{\circ}$ ambiguity is resolved by using the minimum-energy code \citep{Metcalf..1994SoPh..155..235M,Leka..2009SoPh..260...83L}. The inversion provided several physical parameters, including maps of continuum intensity, LOS velocity, magnetic field inclination and so on. The data series have a pixel scale of about 0.5$^\prime$$^\prime$ and a cadence of 12 minutes. These data can show the change of longitudinal and transverse magnetic fields during the evolution of sunspots.

To show the fine structure of the sunspot, we present some TiO images observed by the New Vacuum Solar Telescope (NVST) \citep{Liu..2014RAA....14..705L}. The TiO images were taken with a cadence of 12 s and a pixel size of  0.04$^\prime$$^\prime$. The data were calibrated from Level 0 to Level 1 with the dark current subtracted and flat-field corrected.  To acquired better image quality, the calibrated images were reconstructed with the speckle masking method from Level 1 to Level 1 + \citep{Xiang..2016NewA...49....8X}.

By using the standard procedure in SSW, the continuum intensity and LOS magnetograms data were rotated differentially to a reference time (at 18:00:00 UT on 2017 September 3). The SDO and NVST images were co-aligned by the cross-correlation method.

All the continuum intensity images were normalized to the quiet Sun continuum intensity ($I_0$). The quiet Sun continuum intensity is the average value of continuum intensity in the region of quiet Sun. Each of the continuum intensity images were divided into three regions to precisely identify the umbra and penumbra. The umbra area is defined as the area of continuum image with its intensity darker than 0.48 $I_0$ ( $I_{umbra} \leq 0.48 I_0$ )(\cite{Yang..2018PASP..130j4503Y}). The penumbra area is defined as the area of continuum image with its intensity brighter than 0.48 $I_0$ and darker than 0.85 $I_0$ ( $0.48 I_0<I_{penumbra} \leq 0.85 I_0$ ). The gravity center of sunspot $(X_c,Y_c )$ is defined as:
\begin{equation}\label{equ1}
   X_c =  \frac{\sum I_i x_i}{\sum I_i} \; \textbf{,} \qquad \qquad
   Y_c =  \frac{\sum I_i y_i}{\sum I_i}
\end{equation}

\textbf{The data provided in the $hmi.Sharp$-$cea$-$720s$ were used to product the magnetic quantities (the integrated positive $(\phi_{zp})$ and negative $(\phi_{zn})$ magnetic flux; the transverse magnetic field strength $(B_{t})$ ; and the magnetic field inclination angle $( \gamma )$). The SHARP data are as follows: $B_{P}$, which give the component of the magnetic field, positive westward; $B_{T}$, which give the component of the magnetic field, positive southward; $B_{R}$, which give the radial component of the magnetic field, positive upward and their errors $B_{P-ERR}$, $ B_{T-ERR}$, $ B_{R-ERR}$\citep{Hoeksema..2014SoPh..289.3483H}.}

\textbf{The $B_{P}$, $B_{T}$ and $B_{R}$ in the cylindrical equal area (CEA) coordinate system were conversed to $B_{x}$, $B_{y}$ and $B_{z}$ in plate coordinates ($B_{x}=B_{P}$, $B_{y}= -B_{T}$, $B_{z}=B_{R}$).  The ${\sigma_{B_{x}}}$, ${\sigma_{B_{y}}}$ and ${\sigma_{B_{z}}}$ are the errors of $B_{x}$, $B_{y}$ and $B_{z}$ (${\sigma_{B_{x}}}=B_{P-ERR}$, ${\sigma_{B_{y}}}= B_{T-ERR}$, ${\sigma_{B_{z}}}=B_{R-ERR}$ ).}
The integrated positive $(\phi_{zp})$ and negative $(\phi_{zn})$ magnetic flux in sunspot are according to the following equations:
\begin{equation}\label{equ2}
  {\phi}_{zp} = \int B_{z+} dA \; \textbf{,} \qquad \qquad
  {\phi}_{zn} = \int B_{z-} dA
\end{equation}
where $B_{z+}$ /$B_{z-}$ is the positive/negative longitudinal magnetic field and the $ dA $ denotes the integrated area (sunspot area).

\textbf{The transverse magnetic field strength $(B_{t})$ is calculated by following equation:}
\begin{equation}\label{equ3}
  {B_{t}} = \sqrt{{B_{x}}^{2}+{B_{y}}^{2}}
\end{equation}
\textbf{where the $B_{x}$ and $B_{y}$ is the horizontal components of magnetic field in plate coordinates.}

The magnetic field inclination angle $(\gamma \text{, unit is degree})$ is calculated by following equation:
\begin{equation}\label{equ5}
  {\gamma} = \arctan{(\frac{B_{t}}{\mid B_{z}\mid })} \cdot (\frac{180}{\pi})
\end{equation}
where the $ B_{t} $ is the transverse magnetic field and $B_{z}$ is the longitudinal magnetic field.

\textbf{According to the error propagation, the error of magnetic flux is given by}
\begin{equation}\label{equ4}
  \sigma_{\phi_{z+}} = (\frac{\partial{{\phi}_{z+}}} {\partial{B_{z+}}} ) \cdot {\sigma_{B_{z+}}} = \sum{\sigma_{B_{z+}}} \; \textbf{,} \qquad \qquad
  \sigma_{\phi_{z-}} = ( \frac{\partial{{\phi}_{z-}}} {\partial{B_{z-}}} ) \cdot {\sigma_{B_{z-}}} = \sum{\sigma_{B_{z-}}}
\end{equation}
\textbf{where the ${\sigma_{B_{z+}}}$/${\sigma_{B_{z-}}}$ is the error of the positive/negative longitudinal magnetic field in the sunspot area.}

\textbf{The error of the transverse magnetic field strength $(\sigma_{B_{t}})$ is given by }
\begin{equation}\label{equ4}
  {\sigma_{B_{t}}} = ( \frac{\partial{B_{t}}} {\partial{B_{x}}} ) \cdot {\sigma_{B_{x}}} + (\frac{\partial{B_{t}}} {\partial{B_{y}}} ) \cdot {\sigma_{B_{y}}}
\end{equation}

\textbf{The error of the magnetic field inclination angle $({\sigma_{\gamma}})$ is given by:}
\begin{equation}\label{equ6}
  {\sigma_{\gamma}} = (\frac{\partial{\gamma}} {\partial{B_{z}}} ) \cdot {\sigma_{B_{z}}} + (\frac{\partial{\gamma}} {\partial{B_{t}}}) \cdot {\sigma_{B_{t}}}
\end{equation}
\textbf{where the $\sigma_{B_{z}}$ is the error of the longitudinal magnetic field and $\sigma_{B_{t}}$ is the error of the transverse magnetic field.}
\begin{figure}
\plotone{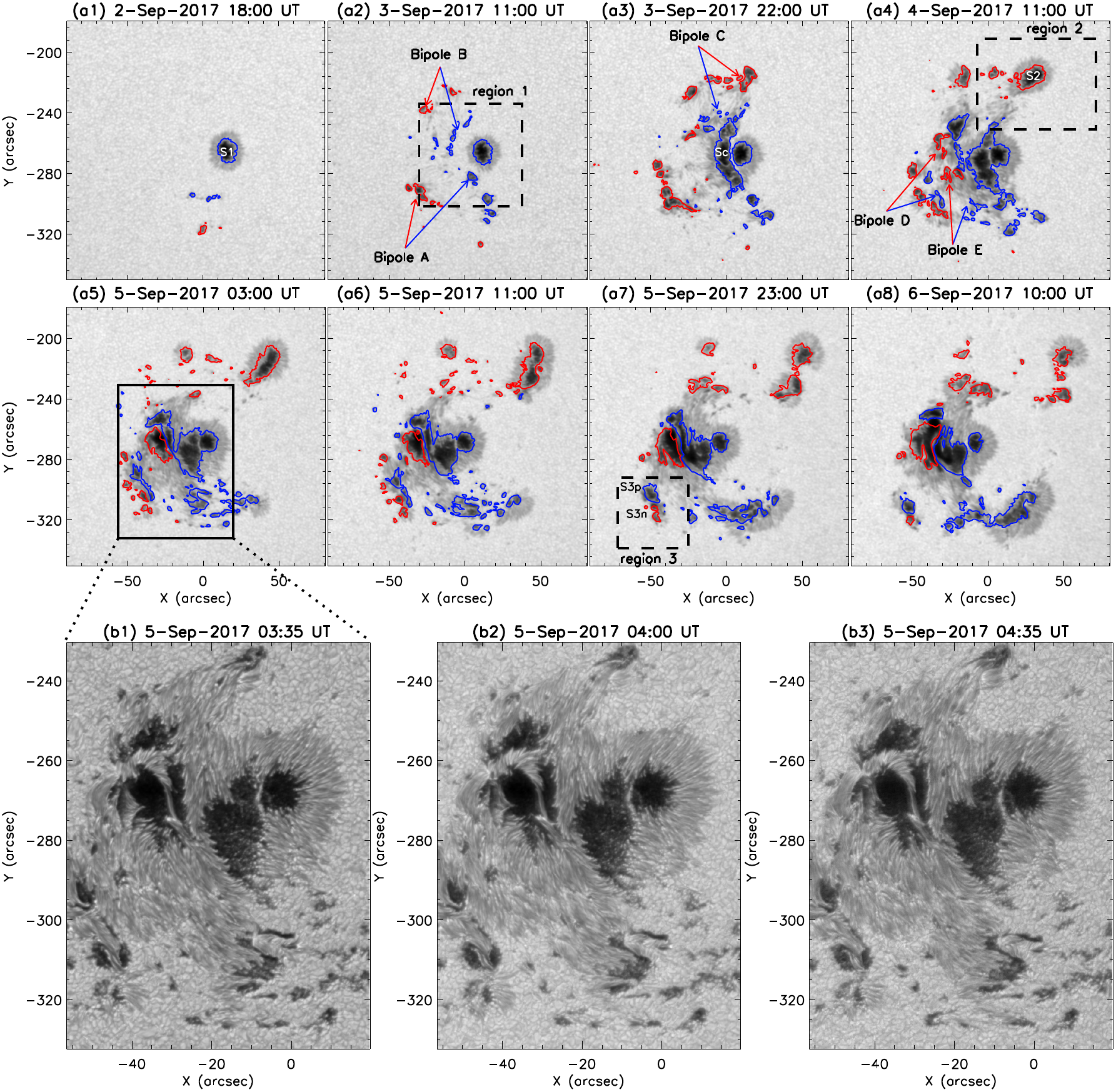}
\caption{Evolution of Active Region NOAA 12673. (a1)-(a8): Continuum intensity images observed by SDO/HMI. The red and blue contours represent the LOS magnetic fields at -800G and +800G, respectively. The boxes (black dashed line) in Figure \ref{fig1} (a2), in Figure \ref{fig1} (a4) and in Figure \ref{fig1} (a7) outline the field of view of Figures \ref{fig2}, Figures \ref{fig8} and Figures \ref{fig12}, respectively. (b1)-(b2): High-resolution TiO images observed by the NVST. \label{fig1}}
\end{figure}

In addition, we traced the photospheric horizontal motions of the plasma flow around the sunspots by the differential affine velocity estimator for vector magnetograms (DAVE4VM) method \citep{Schuck...2008ApJ...683.1134S}. The SDO/AIA 1600 {\AA} images and the GOES X-ray flux profile of 0.1-0.8 nm are utilized to investigate the role of solar flare in penumbral decay.

\section{Results}\label{sec:results}
\subsection{The evolution of Active region NOAA 12673}
Active region (AR) NOAA 12673 appeared at the east solar limb on 2017 August 30 and disappeared at the west solar limb on 2017 September 9. The AR 12673 was classified as a $\alpha$ magnetic field configuration of the sunspot group and consisted of a simple mature sunspot (S1) with a fully extended penumbra from 2017 August 30 to September 2. The sunspot S1 with positive polarity was located on S08 E11 at 18:46 UT on 2017 September 2 (see Figure \ref{fig1}(a1)).

On 2017 September 3, some new flux emerged as bipole (labeled with ``Bipole A" and ``Bipole B") near the sunspot S1. The small patches in the continuum intensity images showed the footpoints of the emerging magnetic flux. The negative patches of the Bipole A and Bipole B moved eastward and the positive ones moved westward (see Figure \ref{fig1}(a2)). The westward patches gradually combined together and then a C-shaped sunspot (Sc) formed (see Figure \ref{fig1}(a3)). In particular, the penumbra of the C-shaped sunspot appeared in the side toward the magnetic flux emergence. As the sunspot Sc gradually got closer to the S1, the penumbra of sunspot S1 nearby the emerging region disappeared. This region was labeled by ``region 1'' (see the black dashed box in Figure \ref{fig1}(a2)).

In next few hours, the AR NOAA 12673 changed rapidly and the magnetic configuration of the sunspot group gradually developed into a $\beta\gamma$-type. The AR 12673 began to erupt a series of flares on September 4. When the AR 12673 went across the visible solar disk, it produced several M-class and X-class flares. Two X-class flares occurred on 2017 September 6. More detailed descriptions of the AR 12673 is given by \cite{Yang..2017ApJ...849L..21Y,Verma..2018A&A...612A.101V,Yan..2018ApJ...856...79Y,Hou..2018A&A...619A.100H,Shen...2018ApJ...861...28S,Jiang...2018ApJ...863..159J,Jiang...2019ApJ...871...16J} and \cite{Jiang...2018ApJ...869...13J}.

Around 22:00 UT on 2017 September 3, another small bipole (labeled with ``Bipole C") emerged around the sunspot S1. The patches of Bipole C with negative polarity gradually moved to the northwestern and coalesced with pre-emerging negative patches of Bipole B (see Figure \ref{fig1}(a3)). These small patches gradually developed into a bigger sunspot (S2) with full penumbra filaments (see Figure \ref{fig1}(a4) and (a5)). After 2017 September 5, the sunspot S2 started to decay and gradually lost its penumbra (see Figure \ref{fig1}(a6)-(a8)). The penumbra of S2 on the side away from the flux emergence disappeared earlier than the other side. This region was labeled by ``region 2'' (see the black dashed box in Figure \ref{fig1}(a4)).

At around 11:00 UT on 2017 September 4, two new bipoles appeared (labeled with ``Bipole D" and ``Bipole E", see Figure \ref{fig1}(a4)). The positive footpoints of  Bipole D gradually moved southward and combined together to form a small positive sunspot (S3p) (see high-resolution observations of TiO images in Figure \ref{fig1}(b1)-(b2)). The southward motion of Bipole D was blocked by the negative sunspot S3n. Then the negative sunspot S3n gradually disappeared and the penumbra of the sunspot S3p began to decay. In the end, the negative sunspot S3n almost completely disappeared and the positive sunspot S3p became a naked pore (see Figure \ref{fig1}(a5)-(a8)). This region was labeled by ``region 3'' (see the block dashed box in Figure \ref{fig1}(a5)).

The following sections discuss the evolution of sunspot penumbra in these three regions. In region 1, we study the formation of penumbra in sunspot Sc and the disappearance of penumbra in sunspot S1. In region 2, we present the evolution of sunspot S2, including its penumbra formation and decay. In region 3, we discuss the penumbra decay of sunspot S3n and S3p. For comparison, we also discuss the photospheric magnetic field properties in umbra during the evolution of these sunspots.

\begin{figure}
\plotone{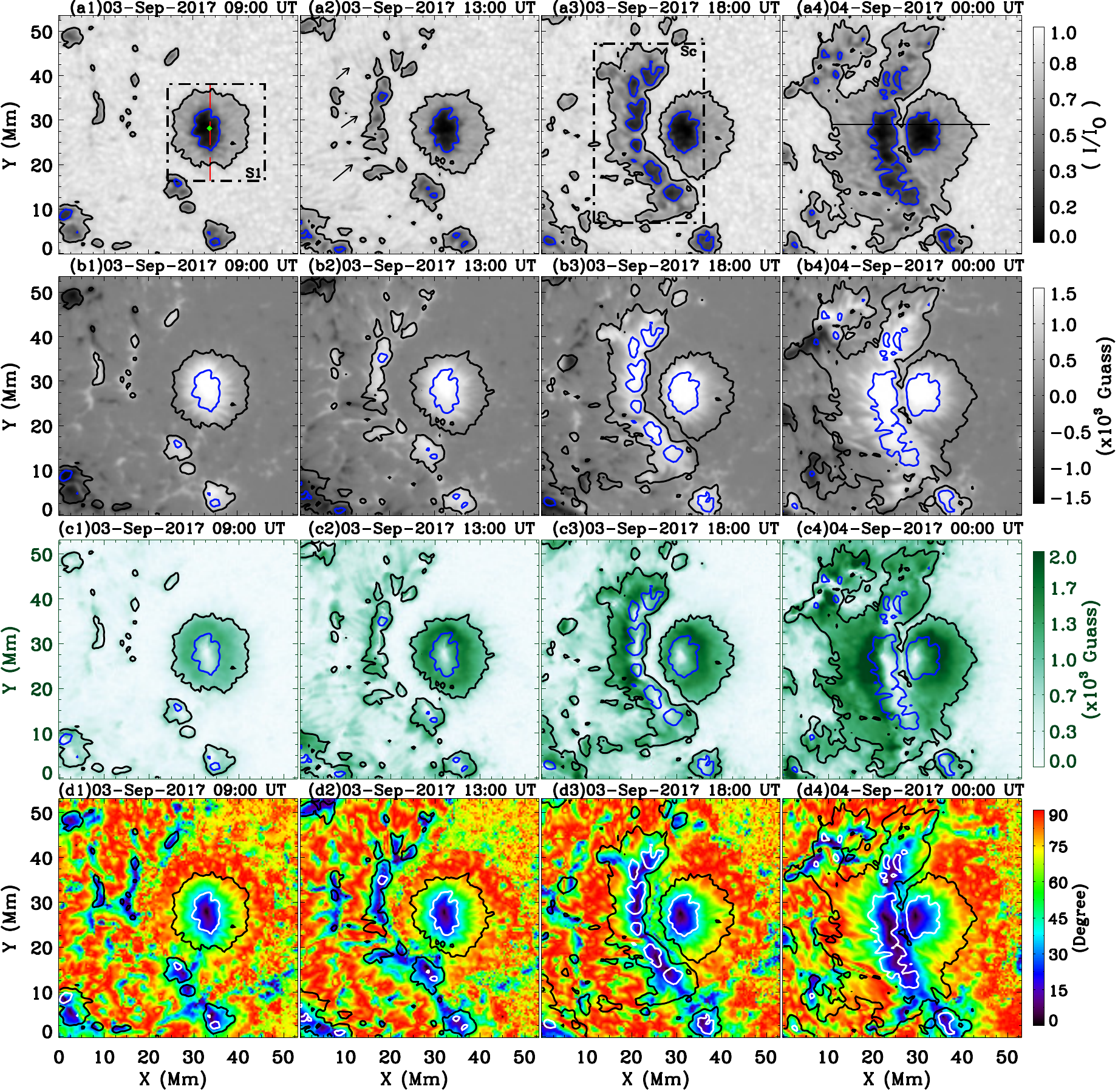}
\caption{Temporal evolution of sunspots S1 and Sc in Region 1 from 09:00 UT on September 3 to 00:00 UT on September 4. (a1)-(a4): Continuum intensity images observed by SDO/HMI. The green dot in Figure 2 (a1) indicates the gravity center of the sunspot S1. The horizontal black line in Figure 2 (a4) indicates the position of the slit for the time-distance diagram in Figure 5 (a). (b1)-(b4): Longitudinal magnetic field maps. (c1)-(c4): Transverse magnetic field maps. (d1)-(d4): Magnetic inclination angle maps. The black and the blue contours represent the boundaries of the penumbra and umbra, respectively. Particularly, the white contours represent the boundaries of the umbra in the last panels((b1)-(b4)). \label{fig2}}
\end{figure}
\subsection{The evolution of sunspots in region 1}
\begin{figure}
\plotone{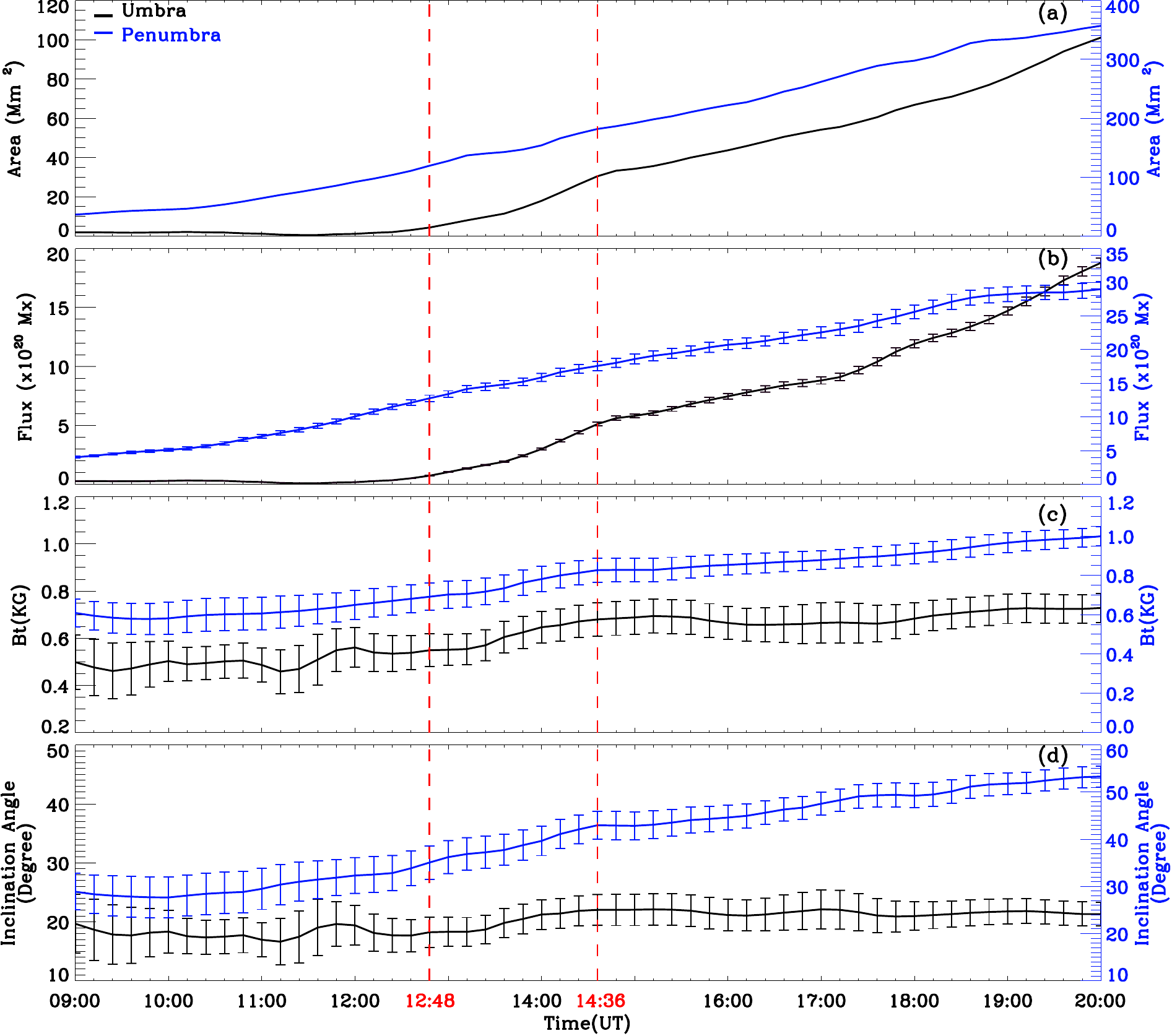}
\caption{Variations of the area (a), the total positive magnetic flux (b), the mean transverse magnetic field strength (c), and the mean magnetic inclination angle (d) inside the umbra and penumbra of Sc from 09:00 UT to 20:00 UT on 2017 September 3. The black and the blue solid curves indicate the parameter variations of the umbra and penumbra, respectively. \label{fig3}}
\end{figure}

Figure \ref{fig2} shows the formation of penumbra in Sc and the disappearance of penumbra in S1 with the continuum intensity images ((a1)-(a4)), longitudinal magnetic field maps ((b1)-(b4)), transverse magnetic field maps ((c1)-(c4)) and magnetic inclination angle maps ((c1)-(c4)).

From the continuum intensity images and the longitudinal magnetic field maps (see in Figure \ref{fig2}(a1)-(a4) and (b1)-(b4)), the patches of Bipoles A and B with the positive polarity merged together and formed a C-shaped sunspot, Sc. Before the appearance of penumbra in Sc, there were some elongated granules along the direction connecting two opposite polarities of Bipoles A and B (see the black arrows in Figure\ref{fig2}(a2)). At around 18:00 UT, the sunspot Sc formed a relatively stable structure, including umbra and penumbra. During this process, the magnetic field strength of the patches gradually increased with time, especially the side towards the sunspot S1. It is noticed that Sc did not have any penumbra on the side towards S1.  When the sunspot Sc gradually approached S1, the left (east) penumbra of S1 gradually disappeared.

From the maps of the transverse magnetic field (see in Figure \ref{fig2}(c1)-(c4)), the transverse magnetic field strength of Sc increased with time during the formation of Sc, especially the side away from S1. The penumbra had a stronger transverse magnetic field than umbra. From the maps of magnetic inclination angle (see in Figure \ref{fig2}(d1)-(d4)), the magnetic inclination angle in umbra was small and changed a little. However, the magnetic inclination angle in penumbra was big and gradually increased with time. During the formation of Sc, more and more horizontal magnetic field appeared in penumbra.

\begin{figure}
\plotone{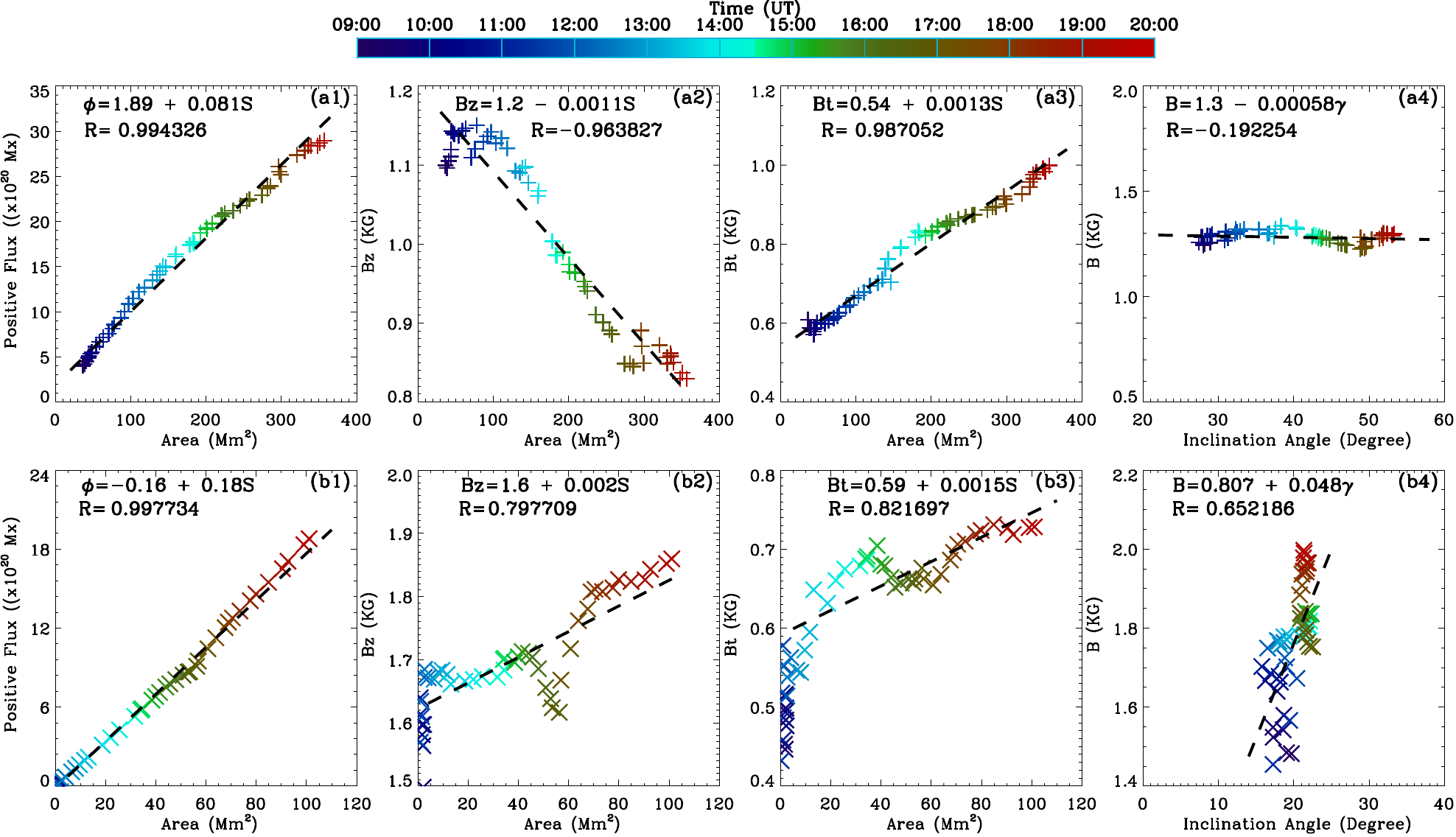}
\caption{Plots of the total magnetic flux, the mean longitudinal magnetic field strength (Bz), and the mean transverse magnetic field strength (Bt) for the penumbra/umbra of Sc as a function of the penumbral area (top, (a1)-(a3))/the umbral area (bottom, (b1)-(b3)) of Sc during the formation of Sc. Plot of the mean magnetic field strength (B) for the penumbra/umbra of Sc as a function of the mean magnetic inclination angle for penumbra (top, (a4))/ umbra (bottom, (b4)) of Sc during the formation of Sc. The dashed lines in the each panel represent the linear fitting results by minimizing the chi-square error statistic. The colours of symbols represent the evolution of time. \label{fig4}}
\end{figure}

Figure \ref{fig3} shows the variations of the area, the total positive magnetic flux, the mean transverse magnetic field strength and the mean magnetic inclination angle in penumbra and umbra during the formation of Sc. The black and the blue solid curves show the variations in umbra and penumbra of Sc, respectively.

From 09:00 UT to 12:48 UT on Sep.3, Sc had a small umbra. After 12:48 UT, the umbra gradually grew in area. During the formation of Sc, the area, the total positive magnetic flux, the mean transverse magnetic field strength and the mean magnetic inclination angle in umbra of Sc increased with time. The mean transverse magnetic field strength in umbra increased from 500G at 12:48 UT to 700G 14:36 UT. In the follow, the value of mean transverse magnetic field strength in umbra remained at around 650-700G. Similarly, the magnetic inclination angle in umbra increased from around $15\,^{\circ}$ at 12:48 UT to $20\,^{\circ}$  at 14:36 UT. Then, the magnetic inclination angle in umbra remained at $20\,^{\circ}$ (see black curves in Figure \ref{fig3}). \textbf{The growth range of mean magnetic inclination angle in umbra is within the errors (the error of the mean magnetic inclination angle is about $2\,^{\circ}$ - $5\,^{\circ}$). The increase is uncertain if it owing to the errors of magnetic inclination angle. The mean magnetic inclination angle in umbra may kept a constant during the formation of penumbra. }

The area, the total positive magnetic flux, the mean transverse magnetic field strength and the mean magnetic inclination angle in penumbra increased with time during the formation of Sc. As the formation of penumbra accompanied by the ongoing emerging flux, the penumbra of Sc did not develop at the expense of umbral magnetic flux. The continuous emerging flux can provide enough magnetic flux for the penumbral formation. During the formation of Sc, the mean transverse magnetic field strength in penumbra increased from about 600G at 09:00 UT to around 800G 14:36 UT, and then slowly increased to above 1000G at 20:00 UT. The mean magnetic inclination angle in penumbra increased from below $30\,^{\circ}$ at 09:00 UT to about $45\,^{\circ}$ at 14:36 UT, and then kept at around $50\,^{\circ}$ (see blue curves in Figure \ref{fig3}). The magnetic field in penumbra became more horizontal during the formation of Sc.

\begin{figure}
\plotone{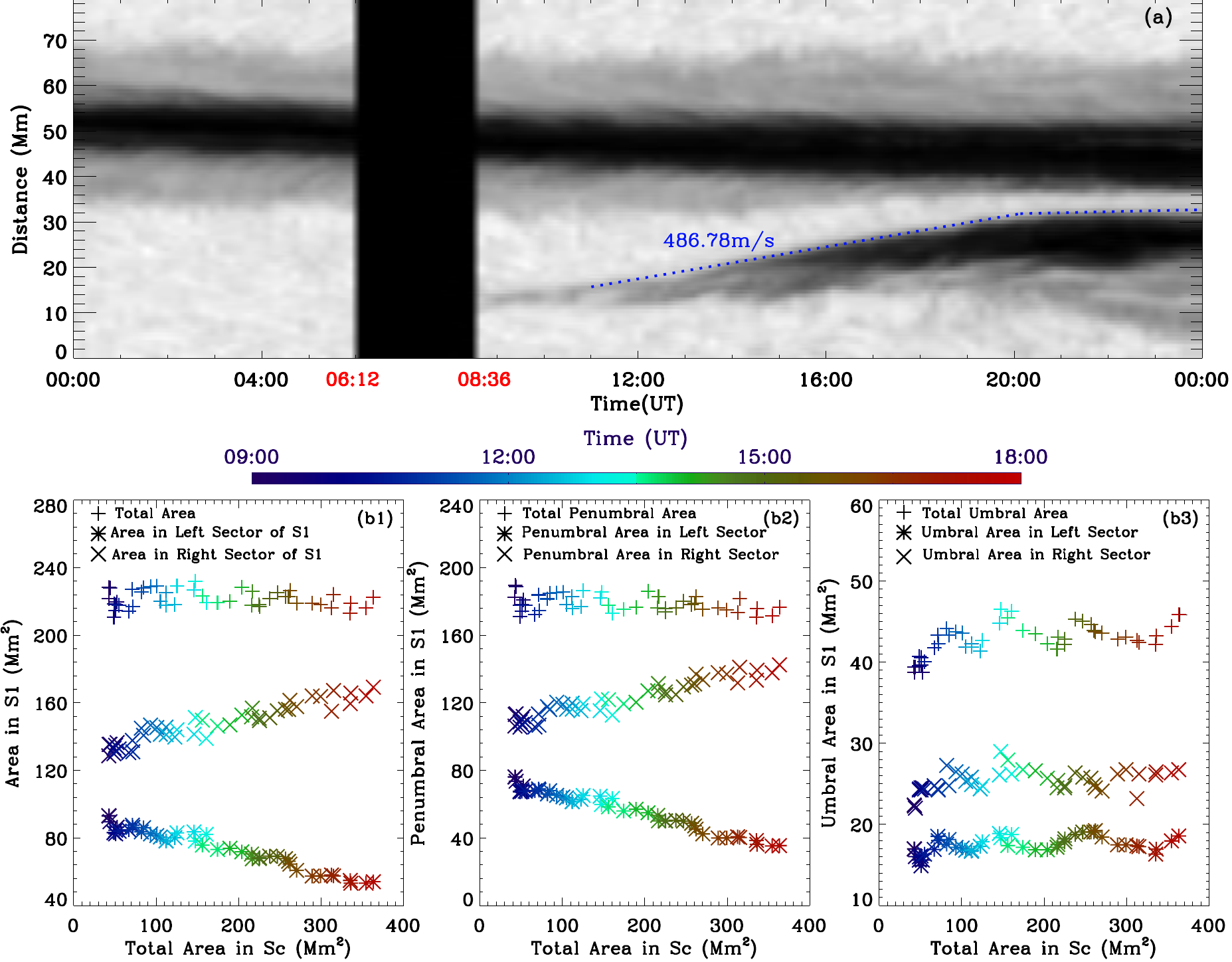}
\caption{(a): Time-distance diagram made by using the continuum intensity images from 00:00 UT on 2017 September 3 to 00:00 UT on 2017 September 4. (b1)-(b3): The area evolution of sunspot S1 during the formation of Sc. The colours of symbol represent the evolution of time. The scatter plots showing the change of the total area (b1)/the penumbral area (b2)/the umbral area (b3) of S1 as a function of the total area of Sc. The ``$+$'' symbols, ``$\ast$'' symbols and ``$\times$'' symbols represent the area in the whole S1, the area in left sector of S1, and the area in the right sector of S1, respectively.
\label{fig5}}
\end{figure}

Figure \ref{fig4} shows that the dependencies of the area and the total magnetic flux/the mean longitudinal magnetic field strength/the mean transverse magnetic field strength, the mean magnetic inclination angle and the mean total magnetic field strength during the formation of Sc. The upper and lower panels in Figure \ref{fig4} show these correlations in penumbra((a1)-(a4)) and umbra((b1)-(b4)), respectively. The dashed lines in the each panel represent the linear fitting results by minimizing the chi-square error statistic. The letter ``R" represents the linear correlation coefficient of two parameters. The colour of symbol represents the evolution of time. The blue and the red symbols respectively indicate the start and end time. Note that the dashed lines, the letter ``R" and the colours in the Figures \ref{fig10}, \ref{fig11} and \ref{fig14} denote the same meaning as Figure \ref{fig4}.

In the penumbra of Sc, the magnetic flux and the mean transverse magnetic field strength increased with the increasing penumbral area (Figure \ref{fig4}(a1) and (a3)). The mean longitudinal field strength (Bz) in the penumbra decreased with the increasing penumbral area (Figure \ref{fig4}(a2)). During the formation of sunspot Sc, the mean total magnetic field strength (B) in penumbra almost keep a constant value with the increasing magnetic inclination angle (Figure \ref{fig4}(a4)). The constant mean magnetic field strength is about 1.3 KG. The increase of the mean magnetic inclination angle in penumbra is from $30\,^{\circ}$ to $55\,^{\circ}$ . These results imply that part of the relatively vertical magnetic lines in the penumbra became more horizontal during the process of penumbral formation.

For the umbra, the magnetic flux, the mean longitudinal/transverse magnetic field strength increased with the increasing umbral area (Figure \ref{fig4}(b1)-(b3)). The mean total magnetic field strength increased with the increasing mean magnetic field inclination in umbra (Figure \ref{fig4}(b4)). The mean total magnetic field strength increased from around 1.5 KG to 1.8KG, and the growth range of mean magnetic inclination angle in umbra is from around $15\,^{\circ}$ to $20\,^{\circ}$. The increase of mean magnetic inclination angle in umbra is small. During the formation of Sc, the dependence of the area and the mean longitudinal field strength in umbra is different from that in penumbra. The growth range of mean magnetic inclination angle in umbra ($5\,^{\circ}$ ) is smaller than that in penumbra ($25\,^{\circ}$). The mean total magnetic field strength in penumbra almost keep a constant value (1.3 KG), however, the mean total magnetic field strength in umbra increased around 300G.

\begin{figure}
\plotone{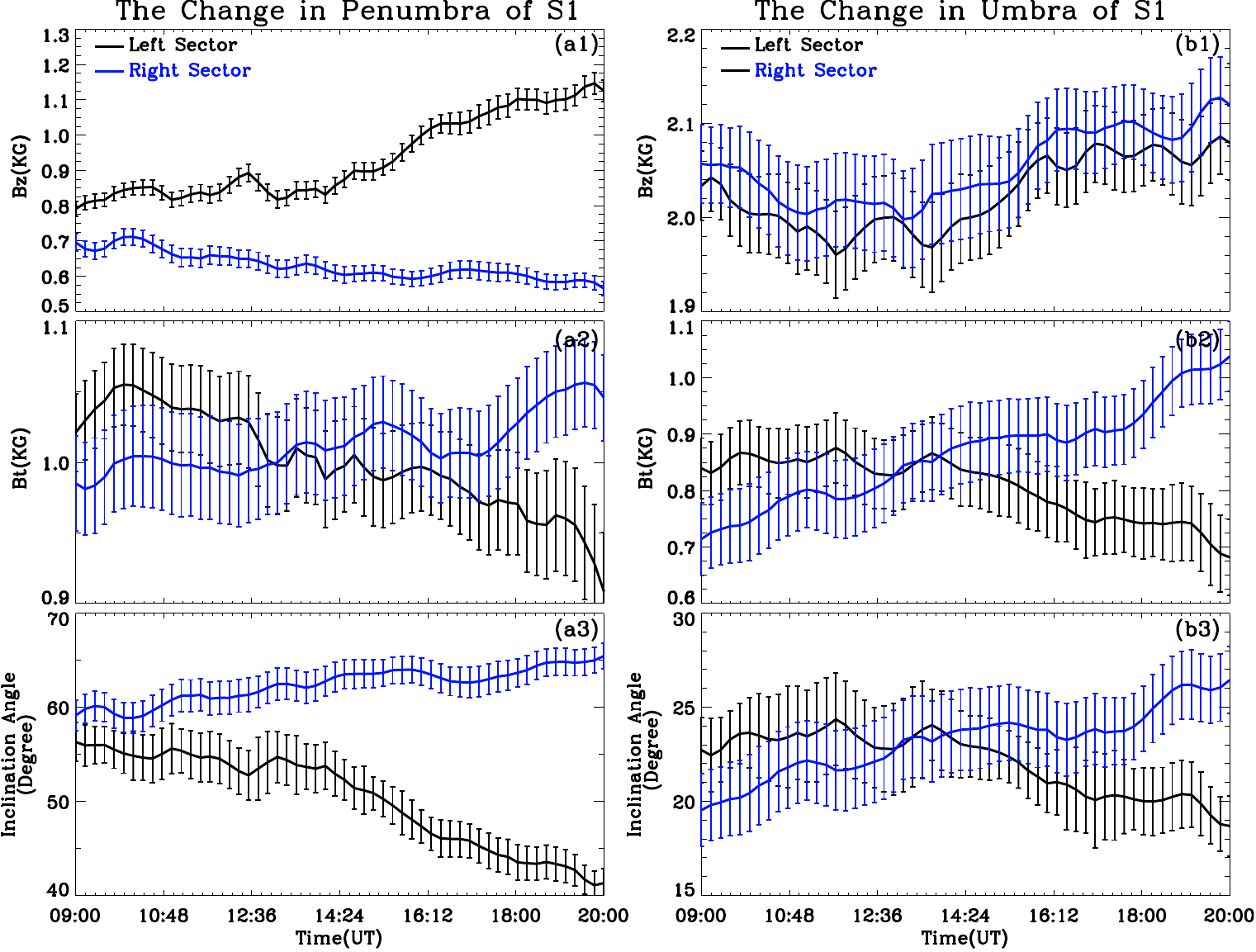}
\caption{Variations of the mean longitudinal magnetic field strength ((a1) and (b1)), the mean transverse magnetic field strength ((a2) and (b2)) and the mean magnetic inclination angle ((a3) and (b3)) in penumbra (solid curves) and umbra (dotted curves) of S1 from 09:00 UT to 20:00 UT on 2017 September 3. The sunspot S1 was divided into the right and left sectors according to the gravity center of S1. \label{fig6}}
\end{figure}

As shown in Figure \ref{fig2}(a1)-(a4), when Sc gradually approached S1, the left (east) penumbra of S1 gradually disappeared. The disappearance of penumbra of S1 has a close relationship with the evolution of Sc. Figure \ref{fig5}(a) shows the time-distance diagram by using the continuum intensity images along the slit across S1 marked by the black line in Figure \ref{fig2}(a4). Seen in the time-distance diagram, Sc approached  S1 at a rate of about 486 m/s during 11:00 UT to 20:00 UT on Sep.3. The left penumbra of S1 started to get shorter at around 13:00 UT. After 18:00 UT, the left penumbra of S1 almost disappeared. Interestingly, the umbra in the left side of S1 and the penumbra in the right side of S1 seem to increase in area.

In order to address the area change of S1 in detail, the sunspot S1 was divided into the left and right sector according to its gravity center, as the red vertical dashed line shown in Figure \ref{fig2}(a1). Figures \ref{fig5}(b1)-(b3) show the relationships between the total area of Sc and the area in each sector of S1. The total area of sunspot is defined as the area in the continuum intensity images darker than 0.85 $I_0$.

Although the total area of  S1 changed little when the total area of Sc increased ( ``$+$'' symbols, in Figure \ref{fig5}(b1)), the sunspot S1 indeed lost a part of its penumbra with the development of Sc. During the formation of Sc, the area in the left sector of S1 decreased (``$\ast$'' symbols) and in the right sector of S1 increased (``$\times$'' symbols) with the increasing total area of Sc. This opposite change in the left and right sector may be the reason why the total area of S1 kept a constant value.
\begin{figure}
\plotone{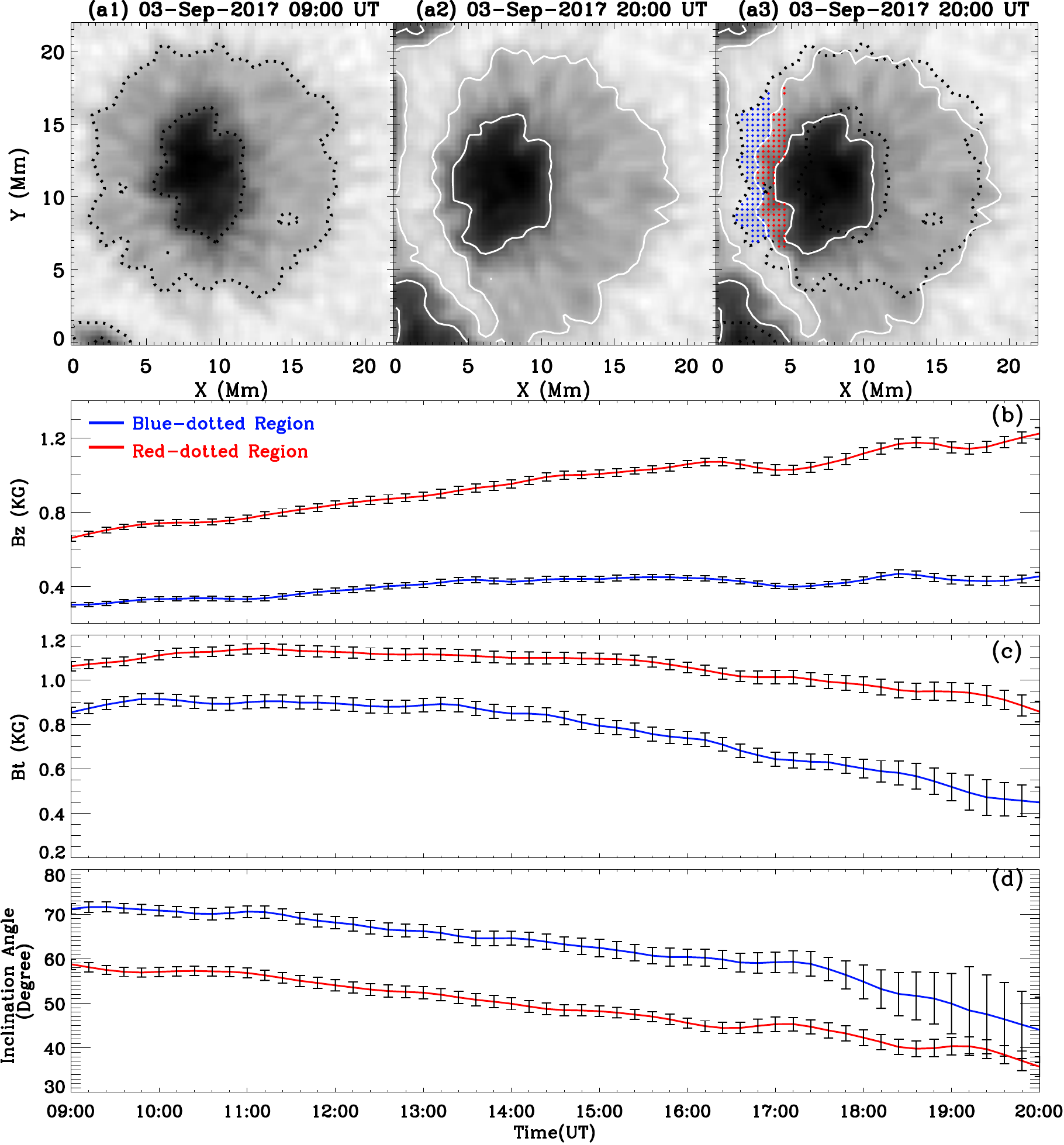}
\caption{(a1)-(a3): The penumbral disappearance of S1 seen in continuum intensity images. The blue-dotted region denotes the region where the penumbra disappeared and the red-dotted region denotes the region where the penumbra still remained in the left sector of S1. (b)-(d): Variation of the mean longitudinal magnetic field strength (b), the mean transverse magnetic field strength (c) and the mean magnetic inclination angle (d) in the blue-dotted region (the blue solid curves) and the red-dotted region (the red solid curves). \label{fig7}}
\end{figure}

Figure \ref{fig5}(b2) shows the relationships between the total area of Sc and the penumbral area in each sector of S1. The total penumbral area of S1 decreased a little when the area of Sc increased ( ``$+$'' symbols). The penumbral area in the left sector decreased (``$\ast$'' symbols) and in the right sector increased (``$\times$'' symbols) with the increasing total area of Sc. Figure \ref{fig5}(b3) shows the relationships between the total area of Sc and the umbral area in each sector of S1. The umbral area in each sector of S1 has no notable change with the increasing total area of Sc (Figure\ref{fig5}(b3)). The total umbral area of S1 ( ``$+$'' symbols), the umbral area in the left sector (``$\ast$'' symbols) and the umbral area in the right sector (``$\times$'' symbols) of S1 increased a little when the area of Sc increased. With the growth of area in Sc, the closer Sc got to S1 and the more penumbral area in the left sector of S1 disappeared. Meanwhile, the more penumbra in the right sector of S1 and the umbra in whole S1 grew.

Figure \ref{fig6} shows the variations of the mean magnetic inclination angle, the mean transverse magnetic field strength, and the mean longitudinal magnetic field strength in penumbra (a1)-(a3) and umbra (b1)-(b3) during the disappearance of penumbra of S1. The black and the blue curves show the variations of the left sector and the right sector of S1, respectively.

\begin{figure}
\plotone{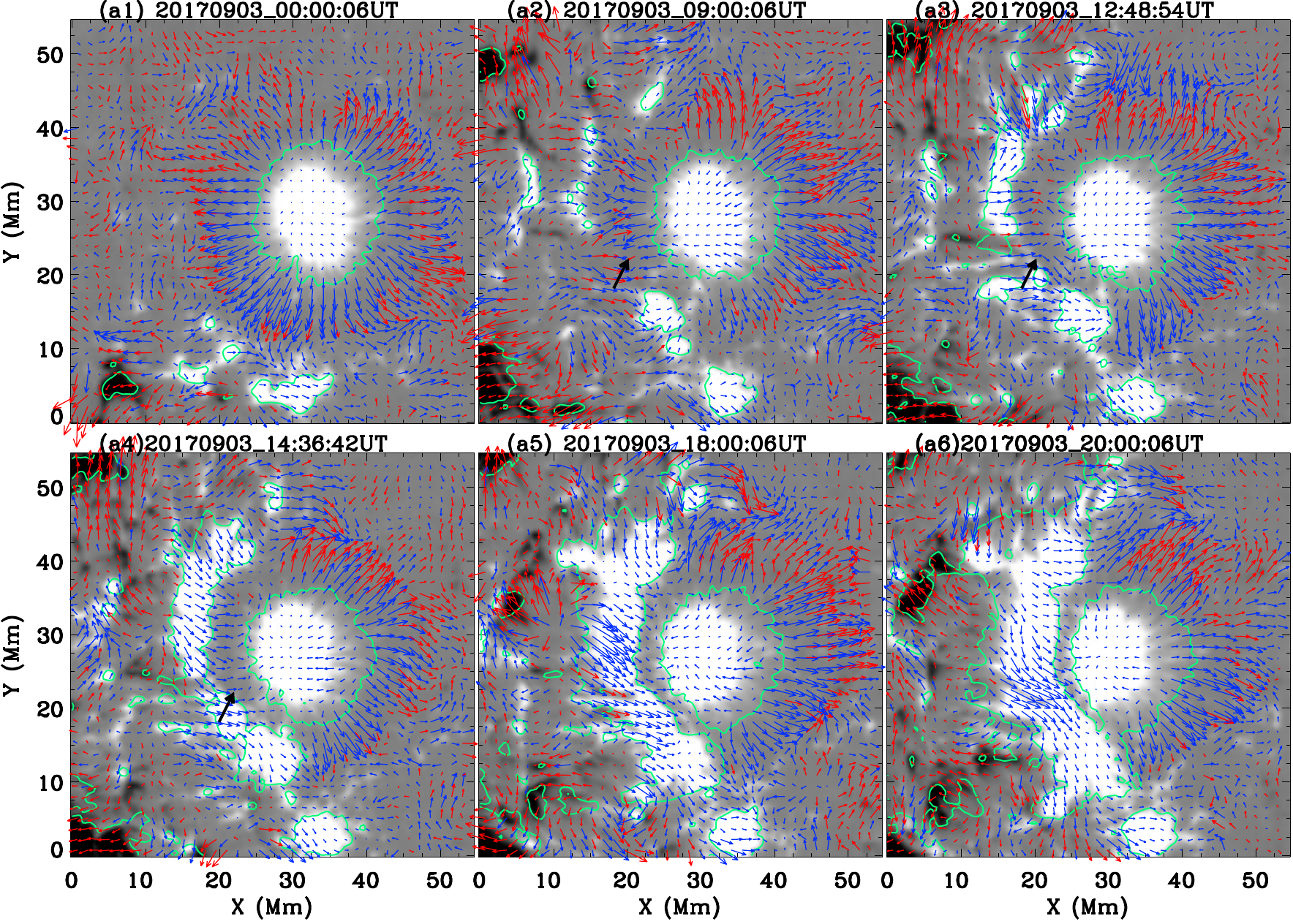}
\caption{Horizontal velocity in the photosphere in Region 1 derived by DAVE4VM. The background image shows the vertical magnetic field with the positive field in white and negative field in black. The arrows represent horizontal velocity. Blue (red) arrows indicate that the vertical magnetic fields in the pixels are positive (negative). The green contours show the sunspot-granulation boundary. The contours shown here and in Figure \ref{fig11b}, \ref{fig14b} are extracted from the continuum intensity images from SDO/HMI. \label{fig7b}}
\end{figure}
In the penumbra of S1, the mean longitudinal magnetic field strength in the left sector increased \textbf{from 800G at 09:00 UT to 1100G at 20:00 UT}, however, it decreased \textbf{from 700G at 09:00 UT to 550G at 20:00 UT} in the right sector (see in Figure \ref{fig6}(a1)). The mean transverse magnetic field strength/the mean magnetic inclination angles in the penumbra of the left sector decreased  \textbf{from around 1000G/$57\,^{\circ}$ at 09:00 UT to 900G/$40\,^{\circ}$ at 20:00 UT}, but it increased \textbf{from around 1000G/$60\,^{\circ}$ at 09:00 UT to 1050G/$65\,^{\circ}$ at 20:00 UT in the penumbra of the right sector} (see in Figure \ref{fig6}(a2) and (a3)). \textbf{Although the increase of the mean transverse magnetic field strength/the mean magnetic inclination angles in the penumbra of the right sector is within their errors, it is notable that the mean longitudinal magnetic field in right sector is decreasing. } The magnetic field of the penumbra became more vertical in left sector and became more horizontal in right sector during the disappearance of penumbra of S1.

In the umbral region, the mean longitudinal magnetic field strength both in the left sector/the right sector increased \textbf{from 2040G/2060G at 09:00 UT to 2080G/2120G at 20:00 UT} (see in Figure \ref{fig6}(b1)). The mean transverse magnetic field strength/the mean magnetic inclination angles decreased \textbf{from 840G/$23\,^{\circ}$ at 09:00 UT to 675G/$19\,^{\circ}$ at 20:00 UT} in the umbra of the left sector, however, it increased \textbf{from 700G/$19\,^{\circ}$ at 09:00 UT to 1025G/$26\,^{\circ}$ at 20:00 UT} in the umbra of the right sector (see in Figure \ref{fig6}(b2) and (b3)). \textbf{Although the variations of the mean magnetic inclination angles and the mean longitudinal magnetic field strength in the umbra are within the errors, the tend of the mean transverse magnetic field strength in umbra is clear. The mean transverse magnetic field strength decreased in left umbra while it increased in right umbra.} In summary, the magnetic field in the left sector of S1 became more vertical and it became more horizontal in the right sector of S1, especially in penumbra. The rearrangement of magnetic field occurred not just in the left sector but in the entire sunspot S1 simultaneously. Furthermore, the rearrangement in the magnetic field of the sunspot S1 caused the disappearance of penumbra in the left sector and the growth of penumbra in the right sector.

Figure \ref{fig7} shows the variations of the mean longitudinal magnetic field strength, the mean transverse magnetic field strength and the mean magnetic inclination angle in the blue-dotted region and the red-dotted region during the disappearance of the left penumbra. The blue-dotted region denotes the area where the penumbra disappeared in the left sector of S1 at 20:00 UT (see the blue dots in Figure \ref{fig7}(a3)). The red-dotted region is the place where the penumbra in the left sector of S1 still remained at 20:00 UT (see the red dots in Figure \ref{fig7}(a3)).

In the blue-dotted region, the mean longitudinal magnetic field strength increased \textbf{from 300G at 09:00 UT to 400G at 20:00 UT,} the mean transverse magnetic field strength/the mean magnetic inclination angle decreased \textbf{from 850G/$70\,^{\circ}$ at 09:00 UT to 450G/$45\,^{\circ}$ at 20:00 UT} (see the blue curves in Figures \ref{fig7}(b)-(d)). In the red-dotted region, the mean longitudinal magnetic field strength increased \textbf{from 700G at 09:00 UT to 1200G at 20:00 UT,}, the mean transverse magnetic field strength/the mean magnetic inclination angle decreased \textbf{from 1050G/$60\,^{\circ}$ at 09:00 UT to 850G/$35\,^{\circ}$ at 20:00 UT} (see the red curves in Figures \ref{fig7}(b)-(d)). The varying trend of the blue-dotted region is consistent with that of the red-dotted region. During the whole process of the penumbral disappearance, the mean magnetic inclination angle in the blue-dotted region is larger than that in the red-dotted region. The strength of the mean transverse magnetic field and the mean vertical magnetic field in the blue-dotted region is weaker than that in red-dotted region. These results indicate that the magnetic field in the left penumbra of the sunspot S1 gradually became more and more vertical from east to west during the disappearance of the left penumbra.

Figure \ref{fig7b} shows the horizontal motion around sunspots (S1 and Sc) derived by DAVE4VM. The prominent flow features around sunspot were a strong outward flow. Before the new flux emerged, the outward flow (moat flow) encircled the sunspot S1 (see Figure \ref{fig7b}(a1)). With the approach of emerging patches, the outward flow in the left side of S1 was diminished. However, the outward motions continued in the other side (see Figure \ref{fig7b}(a2)-(a5)). When the sunspot Sc got closer to the S1, the outward flow in the left side of S1 gradually vanished (see Figure \ref{fig7b}(a6)).

\begin{figure}
\plotone{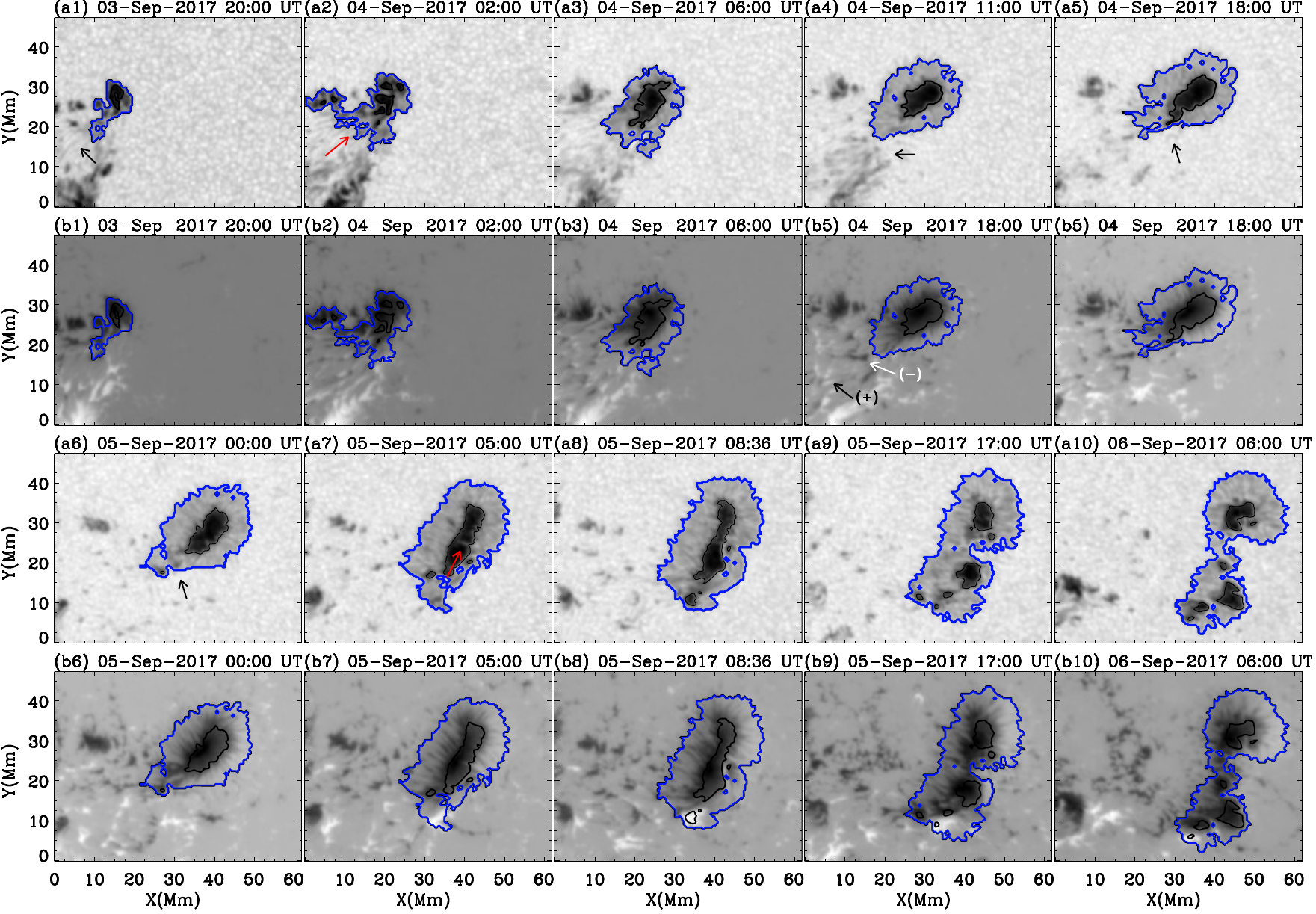}
\caption{Temporal evolution of S2 in the continuum intensity images ((a1)-(a10)) and the LOS magnetograms ((b1)-(b10)) from 20:00 UT on 2017 September 3 to 06:00 UT on September 6. The black and the blue solid curves indicate the boundaries of the umbra and penumbra, respectively.
\label{fig8}}
\end{figure}

\subsection{The evolution of sunspot in region 2}
\begin{figure}
\plotone{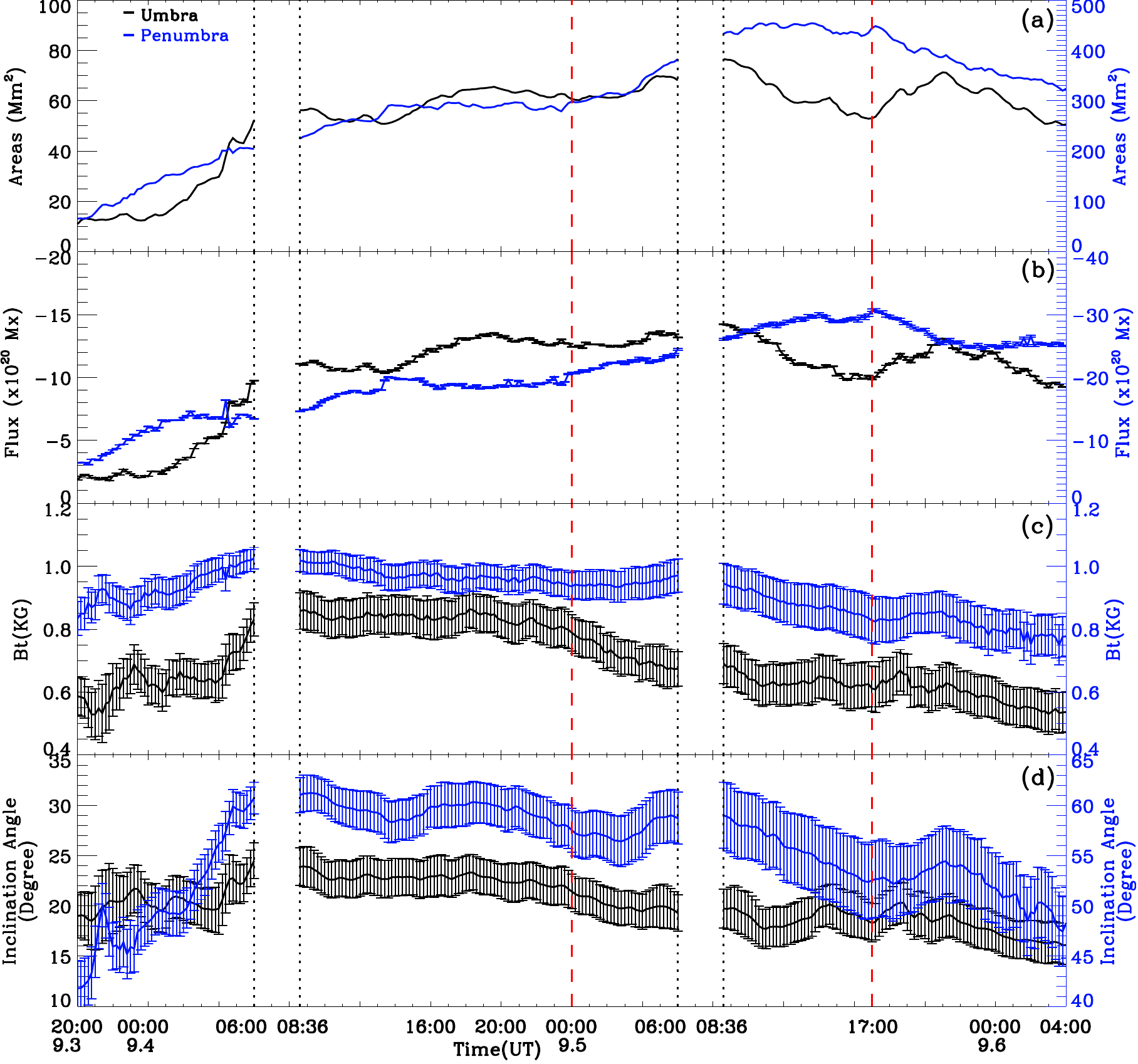}
\caption{Variation of the area (a), the total negative magnetic flux (b), the mean transverse magnetic field strength (c) and the mean magnetic inclination angle (d) in the umbra (black solid curves) and penumbra (blue solid curves) of S2 from 20:00 UT on 2017 September 3 to 04:00 UT on 2017 September 6. The red line indicates the GOES X-ray flux profile of 0.1-0.8 nm in the corresponding time. \label{fig9}}
\end{figure}

Figure \ref{fig8} shows the evolution of sunspot S2 in region 2 in the continuum intensity images and the longitudinal magnetic field maps. The blue and black contours represent the boundaries of penumbra and umbra. SDO observations covered the evolution of S2 from its formation to its decay because S2 evolved quickly.

The formation of S2 originated from the combination of small pores. The combination of small pores was due to the movement of the footpoints of the new emerging magnetic bipoles. These magnetic bipoles connected the negative sunspot S2 and the positive region near the sunspot S1. Some elongated granules were observed between two separating polarities (see the black arrows in Figure \ref{fig8}(a1)). The growth of the sunspot S2 accompanied by the emerging magnetic flux. The penumbra first formed facing the side of flux emergence (see the red arrows in Figure \ref{fig8} (a2)). At around 11:00 UT on Sep.4, a complete sunspot S2 surrounding by a full penumbra had formed. There were still some emerging patches around the complete sunspot S2 (see the black arrows in Figure \ref{fig8}(a4)). The polarity of these patches are opposite (see Figure \ref{fig8}(b4)). The bipolar patches can be interpreted as being produced by sea-serpent field lines \citep{Sainz...2008A&A...481L..21S}. With the approach of these patches, a part of penumbra of S2 disappeared. Then the emerging patches gradually merged with umbra and the shape of umbra changed (see the black arrows in Figure \ref{fig8}(a4)-(a6)).

As a light bridge appeared in the middle of the umbra of S2 at around 05:00 UT on Sep.5, the sunspot S2 started to decay (see the red arrows in Figure \ref{fig8} (a7)). S2 was divided into two parts completely (north part and south part) at 17:00 UT on Sep.5 (see Figure \ref{fig8}(a9)). Then, the south part and the north part of S2 gradually lost their penumbra. The penumbra of S2 on the opposite side of the flux emergence disappeared first (see Figure \ref{fig8}(a10)).
\begin{figure}
\plotone{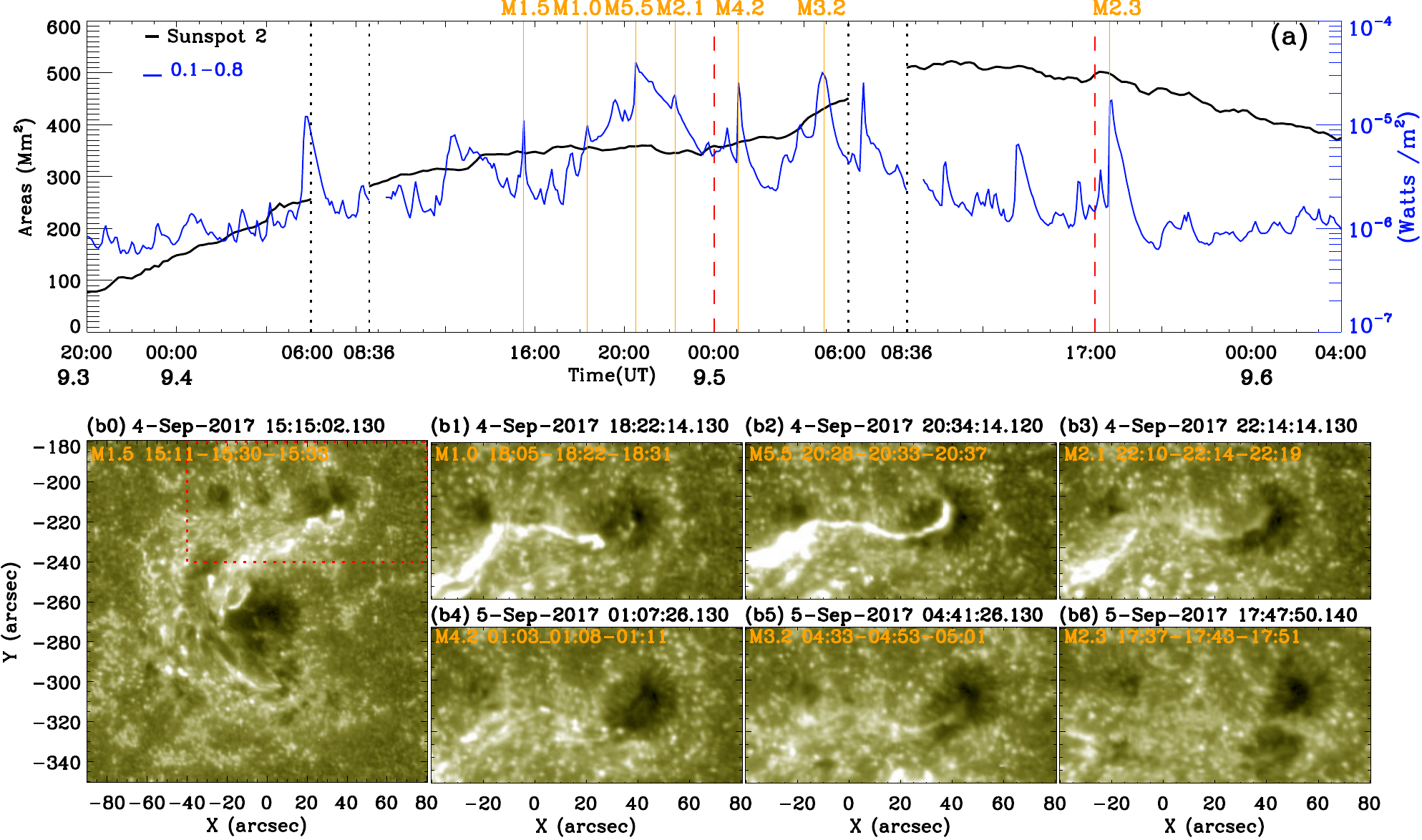}
\caption{(a):The black curve shows the variation of the total area in sunspot S2 from 20:00 UT on 2017 September 3 to 04:00 UT on 2017 September 6. The blue curve indicates the GOES X-ray flux profile of 0.1-0.8 nm in the corresponding time. (b0)-(b6): Temporal evolution of the sunspot S2 in the SDO/AIA 1600{\AA} images. \textbf{The red dashed box in panel (b0) outlines the field of Region 2 (the view of panels (b1)-(b6)).}  } \label{fig9b}
\end{figure}

Figure \ref{fig9} shows the variations of the area, the total negative magnetic flux, the mean transverse magnetic field strength and the mean magnetic inclination angle in umbra (black solid curves) and penumbra (blue solid curves) during the evolution of sunspot S2.

The formation phase of S2 was from 20:00 UT on Sep.3 to 06:00 UT on Sep.5. According to the growing trend of penumbral area, the penumbral formation process can be divided into two stages, the rapidly growing stage and the relatively slow development stage. At the rapidly growing stage, the area in penumbra increased from $70\, Mm^2$ at 20:00 UT on Sep.3 to around $200\, Mm^2$ at 06:00 UT on Sep.4 ($170\, Mm^2$ in 10 hours). At the slowly development stage, the area in penumbra increased from $200\, Mm^2$ at 06:00 UT on Sep.4 to around $380\, Mm^2$ at 06:00 UT on Sep.5 ($180\, Mm^2$ in 24 hours).

At the rapidly growing stage (from 20:00 UT on Sep.3 to 06:00 UT on Sep.4), in the region of umbra, the area, the magnetic flux, the mean transverse magnetic field strength and the mean magnetic inclination angle rapidly increased with time (see the black curves in Figure \ref{fig9}(a)-(d)). Similarly, the area, the magnetic flux, the mean transverse magnetic field strength and the mean magnetic inclination angle in penumbra also rapidly increased (see the blue curves in Figure \ref{fig9}(a) and (d)).

At the slow development stage (from 08:36 UT on Sep.4 to 06:00 UT on Sep.5), the area, the magnetic flux, the mean transverse magnetic field strength and the mean magnetic inclination angle in umbra and penumbra changed with a relatively slow rate, especially the mean transverse magnetic field strength and the mean magnetic inclination angle. During the slow development stage, the mean transverse magnetic field strength in penumbra remained at around 900G - 1000G and the mean magnetic inclination angle in penumbra remained at around $56\,^{\circ}-60\,^{\circ}$. The mean transverse magnetic field strength in umbra remained at around 800G and the mean magnetic inclination angle in umbra remained at around $23\,^{\circ}$ from 08:36 UT on Sep.4 to around 00:00 UT on Sep.5. Then the mean transverse magnetic field strength and the mean magnetic inclination angle in umbra decreased from 800G at 00:00 UT to 650G at 06:00 UT on Sep.5.

The decay phase of the sunspot S2 was from 08:36 UT on Sep.5 to 04:00 UT on Sep.6. From 08:36 UT to 17:00 UT on Sep.5, the area and total negative magnetic flux of the umbra decreased with time. However, at the same time, the area and total negative magnetic flux in penumbra continued to increased. During this period, the flux emergence was not as active as the period of penumbral formation. The penumbra may develop at expense of the umbral magnetic flux. When the umbra completely split into two parts at 17:00 UT, the area and total negative magnetic flux of umbra picked up. In the meantime, the area and total negative magnetic flux of penumbra began to decrease. The umbra may recover by costing the penumbral magnetic flux. The mean transverse magnetic field strength and the mean magnetic inclination angle both in umbra and penumbra gradually decreased during the whole decay phase. The magnetic field of umbra and penumbra became more horizontal during the decay of S2.

Figure \ref{fig9b} simply shows that the relationship between the evolution of S2 and the flare in active region. The blue curve in Figure \ref{fig9b}(a) indicates the GOES X-ray flux profile of 0.1-0.8nm from 20:00 UT on Sep.3 to 04:00 UT on Sep.6. The information of solar flares can be extracted from the GOES flare catalog. The black curve shows the evolution of total area in S2. \textbf{Although a series of flares occurred in the AR NOAA 12673 during the evolution of S2, the evolution of total area in S2 had no abrupt change after the flares. Some brightening can be found in the vicinity of S2 during flares, like the M5.5 flare (see \ref{fig9b}(b2)), the flares may affect the magnetic field in the vicinity of S2. Because of the active surroundings, the sunspot can not be stable. The lifetime of S2 is obvious shorter than the typical mature sunspot.} It typically takes hours to days for sunspot formation, while sunspot decay takes longer, typically weeks to months \citep{Hathaway...2008SoPh..250..269H}.

\begin{figure}
\plotone{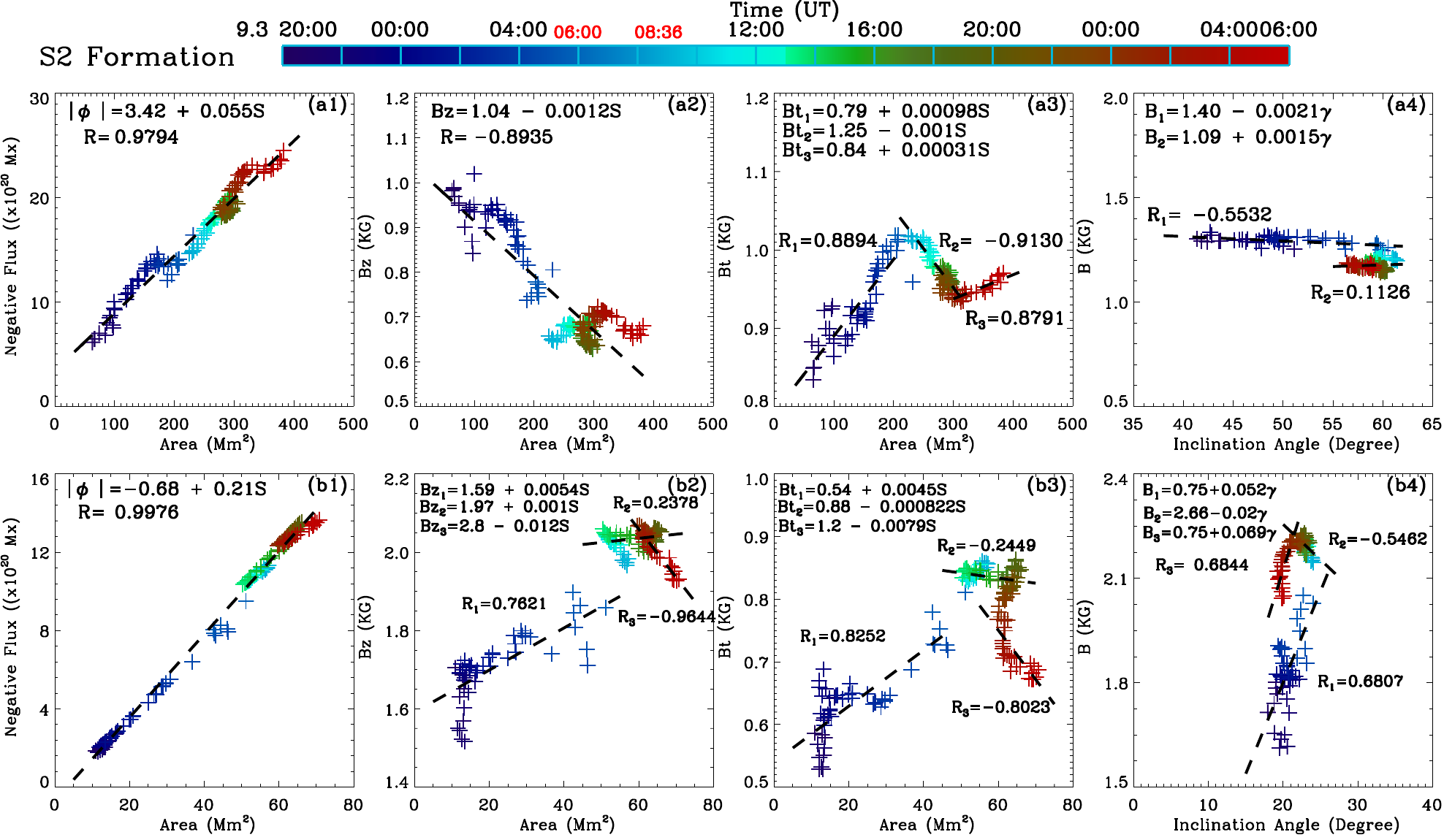}
\caption{Plots of the total magnetic flux, the mean longitudinal magnetic field strength (Bz) and the mean transverse magnetic field strength (Bt) for the penumbra/umbra of the sunspot S2 as a function of the penumbral area (top, (a1)-(a3))/the umbral area (bottom, (b1)-(b3)) of sunspot S2 during the formation of S2. Plot of magnetic field strength (B) for the penumbra/umbra of the sunspot S2 as a function of the mean magnetic inclination angle for penumbra (top, (a4))/umbra (bottom, (b4)) of sunspot Sc during the formation of S2. The colours of symbols represent the evolution of time.
\label{fig10}}
\end{figure}

Figure \ref{fig10} shows that the correlations of the area and the total magnetic flux/the mean longitudinal magnetic field strength (Bz)/the mean transverse magnetic field strength (Bt), the mean magnetic inclination angle and the mean total magnetic field strength during the formation of S2. The upper panels((a1)-(a4)) and lower panels((b1)-(b4)) in Figure \ref{fig10} show these correlations in penumbra and umbra of S2, respectively.

In the penumbra of S2, the magnetic flux increased with the increasing penumbral area (Figure \ref{fig10}(a1)). The mean longitudinal field strength (Bz) in the penumbra decreased with increasing penumbral area (Figure \ref{fig10}(a2)). The relationship between the mean transverse magnetic field strength and the area in the penumbra was divided into three stage: the first one is from 20:00 UT on Sep.3 to 06:00 UT on Sep.4, the rapidly growing stage of S2. The second one is from 08:36 UT on Sep.4 to 00:00 UT on Sep.5. The third one is from 00:00 UT on Sep.5 to 06:00 UT on Sep.5. The mean transverse magnetic field strength increased with the increasing penumbral area during the first stage, decreased during second stage, and increased during the third stage. (Figure \ref{fig10}(a3)). This transition of the tendency in transverse magnetic field may be influenced by the emerging intensity of magnetic flux. At initial stage, a lot of new flux emerged nearby the sunspot S2 and the transverse magnetic field in S2 increased rapidly. However, in the following, the new emerging magnetic flux became less and the transverse magnetic field in S2 started to decrease. At around 00:00 UT on Sep.5, some emerged pathes merged with S2 then the mean transverse magnetic field strength increased.  The relationship between the mean total magnetic field strength (B) and the magnetic inclination angle in the penumbra was divided into two stages: the first one is from 20:00 UT on Sep.3 to 06:00 UT on Sep.4, the rapidly growing stage of S2. The second one is from 08:36 UT on Sep.4 to 06:00 UT on Sep.5, the slow development stage. The mean total magnetic field strength in penumbra decreased a little with the increasing magnetic inclination angle at the rapidly growing stage of S2. And it kept at around 1.09 KG and the magnetic inclination also keep around $55\,^{\circ}-60\,^{\circ}$ during the slow development stage (Figure \ref{fig10}(a4)).
\begin{figure}
\plotone{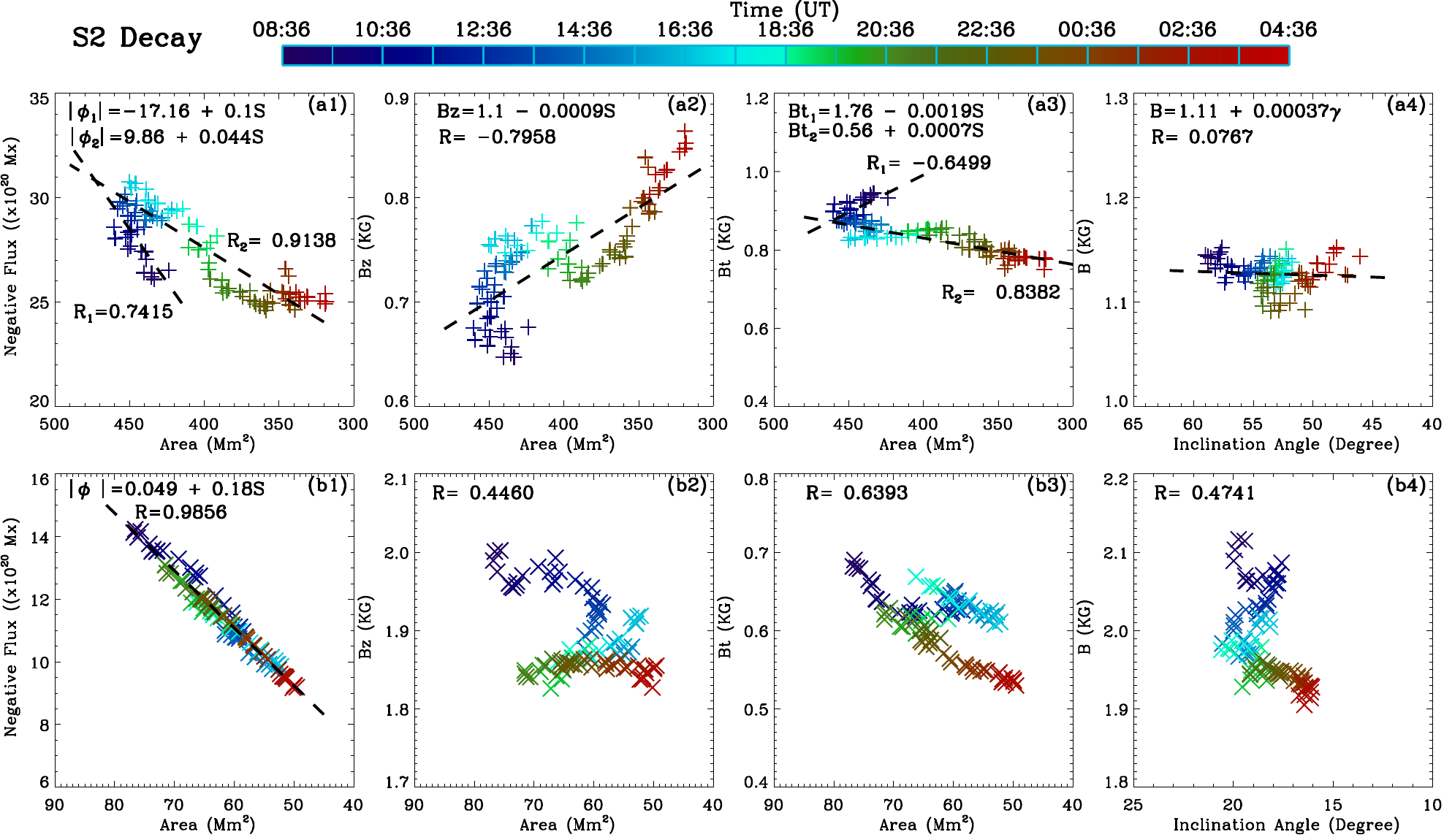}
\caption{Plots of the total magnetic flux, the mean longitudinal magnetic field strength (Bz) and the mean transverse magnetic field strength (Bt) for the penumbra/umbra of the sunspot S2 as a function of the penumbral area (top, (a1)-(a3))/the umbral area (bottom, (b1)-(b3)) of sunspot S2 during the decay of S2. Plot of magnetic field strength (B) for the penumbra/umbra of the sunspot S2 as a function of the mean magnetic inclination angle for penumbra (top, (a4))/umbra (bottom, (b4)) of sunspot S2 during the decay of S2. The colours of symbols represent the evolution of time.\label{fig11}}
\end{figure}

In the umbra of S2, the magnetic flux increased with the increasing umbral area (Figure \ref{fig10}(b1)). The time-division in Figure \ref{fig10}(b2)-(b4) is same as that in Figure \ref{fig10}(a3). The mean longitudinal field strength in umbra increased with the increasing umbral area at the first stage, remained constant (1.97 KG) at the second stage, and decreased at the third stage (Figure \ref{fig10}(b2)). Similarly, the mean transverse magnetic field strength in umbra increased with the increasing umbral area at the first stage, remained constant (0.88 KG) at the second stage, and decreased at the third stage (Figure \ref{fig10}(b3)). It was the same as the dependence of the penumbral area and the mean transverse magnetic field strength in the penumbra of S2, except the third stage. The mean total magnetic field strength in umbra increased with the increasing mean magnetic inclination angle of umbra at the first stage and decreased at the third stage (Figure \ref{fig10}(b4)).

Figure \ref{fig11} shows that the relationships between the area and the total magnetic flux/the mean longitudinal magnetic field strength (Bz)/the mean transverse magnetic field strength (Bt), the mean magnetic inclination angle and the mean total magnetic field strength during the decay of S2. The upper panels((a1)-(a4)) and lower panels((b1)-(b4)) in Figure \ref{fig11} show these relationships in penumbra and umbra of the sunspot S2, respectively. It is worth noting that the area of S2 decreased during its decay and the values of the X-axis in Figure \ref{fig11} is going from big to small.
\begin{figure}
\plotone{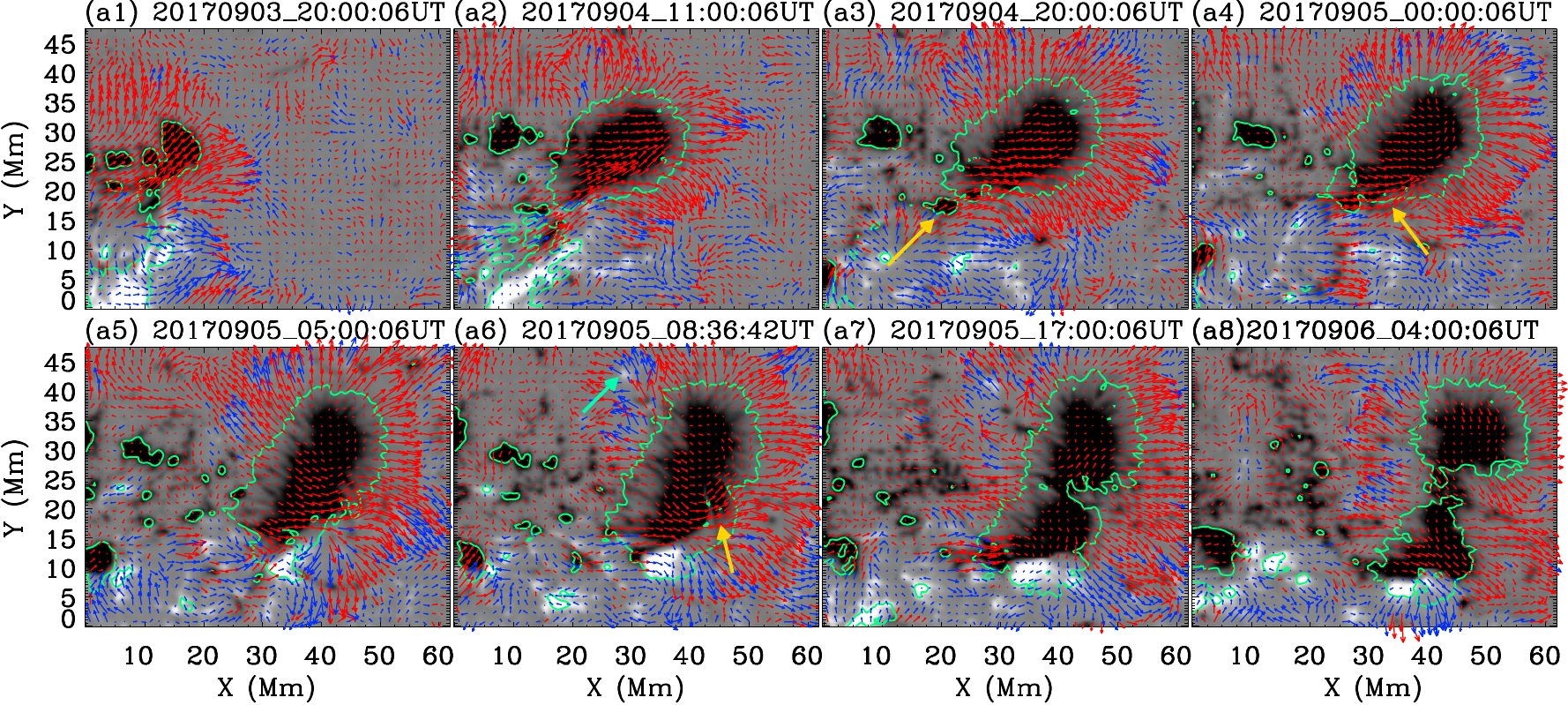}
\caption{ Horizontal velocity in the photosphere in Region 2 derived by DAVE4VM. The background image shows the vertical magnetic field with the positive field in white and negative field in black. The arrows represent horizontal velocity. Blue (red) arrows indicate that the vertical magnetic fields in the pixels are positive (negative).  \label{fig11b}}
\end{figure}

In the penumbra of S2, the relationship between the magnetic flux and the area in the penumbra was divided into two stages: the first one is from 08:36 UT on Sep.5 to 14:00 UT on Sep.5 and the second one is from 14:00 UT on Sep.5 to 04:00 UT on Sep.6. The magnetic flux increased with the increasing penumbral area at the first decay stage, and decreased with the decreasing penumbral area at the second decay stage (Figure \ref{fig11}(a1)). Although the area of umbra in S2 began to decrease at the first stage, the area of penumbra of S2 still increased.  The mean longitudinal field strength in the penumbra gradually increased with the decreasing penumbral area (Figure \ref{fig11}(a2)).  Same in Figure \ref{fig11}(a1), the mean transverse magnetic field strength in the penumbra increased with the increasing penumbral area at the first stage. Then it decreased with the decreasing penumbral area at the second stage (Figure \ref{fig11}(a3)). During the decay of sunspot S2, the mean total magnetic field strength (B) in penumbra changed a little with the decreasing magnetic field inclination (Figure \ref{fig11}(a4)). The mean magnetic field strength kept at 1.1 KG. During the decay of penumbra, part of the horizontal magnetic lines in the penumbra became more vertical.

\begin{figure}
\plotone{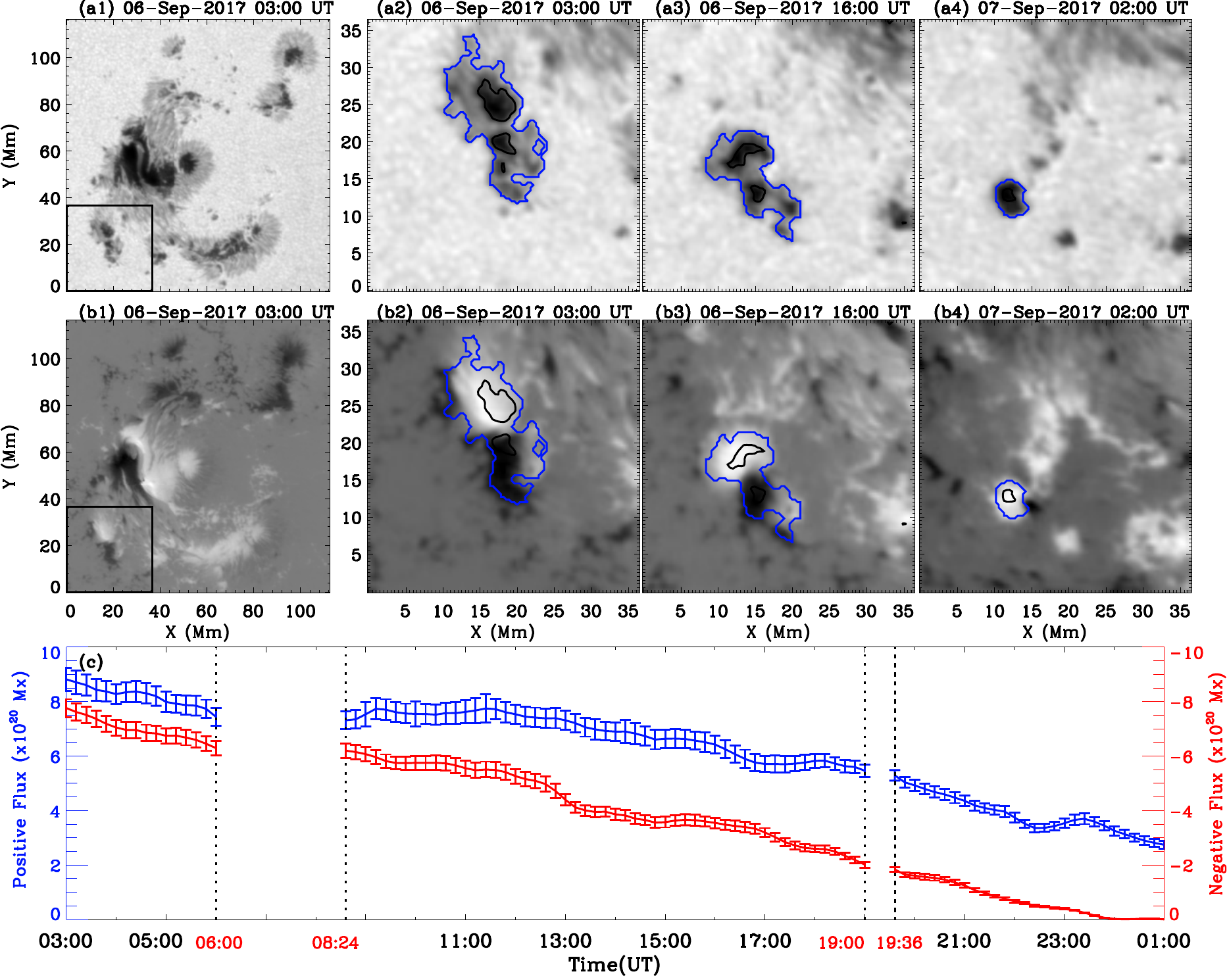}
\caption{The decay process of sunspots in the region 3 from 03:00 UT on 2017 September 6  to 02:00 UT on September 7. (a1)-(a4): Continuum intensity images observed by SDO/HMI. The blue and the black contours represent the boundaries of the penumbra and umbra, respectively. (b1)-(b4): The corresponding LOS magnetograms. (c): Evolution of magnetic flux during the decay of sunspots in the region 3. The red and the blue curves indicate the change of the positive and the negative magnetic flux in the sunspot region where the continuum intensity $I_{sunspot}\leq 0.85 I_0$.  \label{fig12}}
\end{figure}

During the decay of S2, the magnetic flux in umbra decreased with the decreasing umbral area  (Figure \ref{fig11}(b1)). The relationship between the umbral area and the mean longitudinal/transverse magnetic field strength and between the mean total magnetic field strength and the mean magnetic inclination angle of umbra are not notable. Their correlation coefficient is small (Figure \ref{fig11}(b2)-(b4)).

Figures \ref{fig11b} (a1)-(a5)) show the variety of the horizontal motion around S2 during the formation of S2. The sunspot S2 formed in flux emergence. When the S2 was a pore (the penumbra of S2 did not formed), the moat flow of S2 was absent (see Figure \ref{fig11b}(a1)). There was a strong flow in S2, which main caused by the strong flux emergence. During the formation of S2, the outward flow of S2 gradually appeared at the nearest sunspot-granular boundary and its velocity was gradually increased (see Figure \ref{fig11b}(a1)-(a5)). When the S2 was basically formed, the outward flow partly encircled S2. However, the outward flow was absent in the flux emergence (see the yellow arrow in Figure \ref{fig11b}(a3)). With the development of S2, the outward flow did not appear at the side facing the emerged flux. In addition, the typical line, which demarcated radial inflows in the inner and outflow in the outer penumbra \citep{Sobotka..1999A&A...348..621S}, was absent in that side. The strong emerging flow was dominant in this side of S2 (see the yellow arrow in Figure \ref{fig11b}(a4)).  Figure \ref{fig11b} (a6)-(a8)) show the horizontal motion around S2 during the decay of S2. During the decay of S2, the dividing line between inward and outward proper motions in the penumbra \citep{Deng..2007ApJ...671.1013D} was not found either in the side of flux emergence (see the yellow arrow in Figure \ref{fig11b}(a6)). The new emerging flux destroyed the moat flow around the S2 (see the green arrow in Figure \ref{fig11b}(a6)). On the other side away from the flux emergence, the outward flow seem to continue but the flow speeds was lower(see Figure \ref{fig11b}(a6)-(a8)).

\subsection{The decay of sunspot penumbra in region 3}
\begin{figure}
\plotone{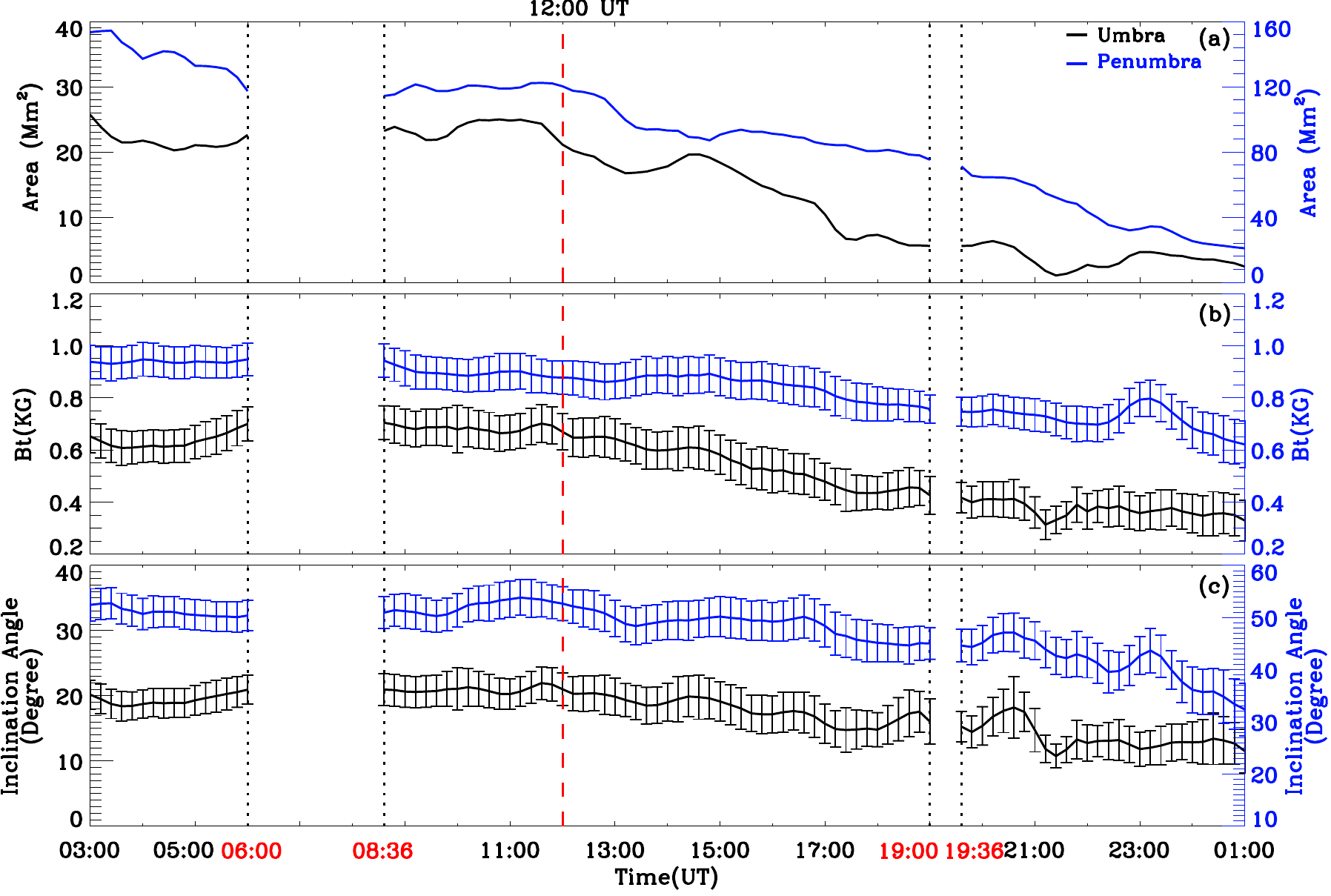}
\caption{Variation of the area (a), the mean transverse magnetic field strength (b) and the mean magnetic inclination angle (c) inside umbra (the black solid curves) and the penumbra (the blue solid curves) of the sunspots in region 3 from 03:00 UT on 2017 September 6  to 01:00 UT on September 7.  \label{fig13}}
\end{figure}

Figure \ref{fig12} shows the decay process of sunspots in region 3 from 03:00 UT on September 6 to 02:00 UT on September 7. Figures \ref{fig12} (a1) and (b1) show the location of region 3 in the AR NOAA 12673. There were two opposite polarity developing sunspots (S3p and S3n) in region 3 (see in Figure \ref{fig12}(a2) and (b2)). When the positive patches of Bipole D moved toward the pre-existing negative one, S3n , the negative sunspot S3n lost its penumbra and disappeared (see in Figure \ref{fig1}(a4)-(a8)). The positive sunspot S3p eventually became a small pore (see in Figure \ref{fig12}(a3) and (b3)). During this process, the positive and negative magnetic flux in the sunspot group (S3p and S3n) decreased with time (see in Figure \ref{fig12}(c)). The result implies that the decay of this sunspot group was accompanied by magnetic cancellation.

Figure \ref{fig13} shows the variations of the area, the mean transverse magnetic field strength and the mean magnetic inclination angle in umbra (the black solid curves) and penumbra (the blue solid curves) of the sunspot group during the decay of sunspot group. Because it is difficult to distinguish the boundary between the positive and negative sunspots in continuum intensity images, the umbral area represents the total area of the two sunspots umbra. The penumbral area is the same as the umbra area, but for the total area of two sunspots penumbra.

\begin{figure}
\plotone{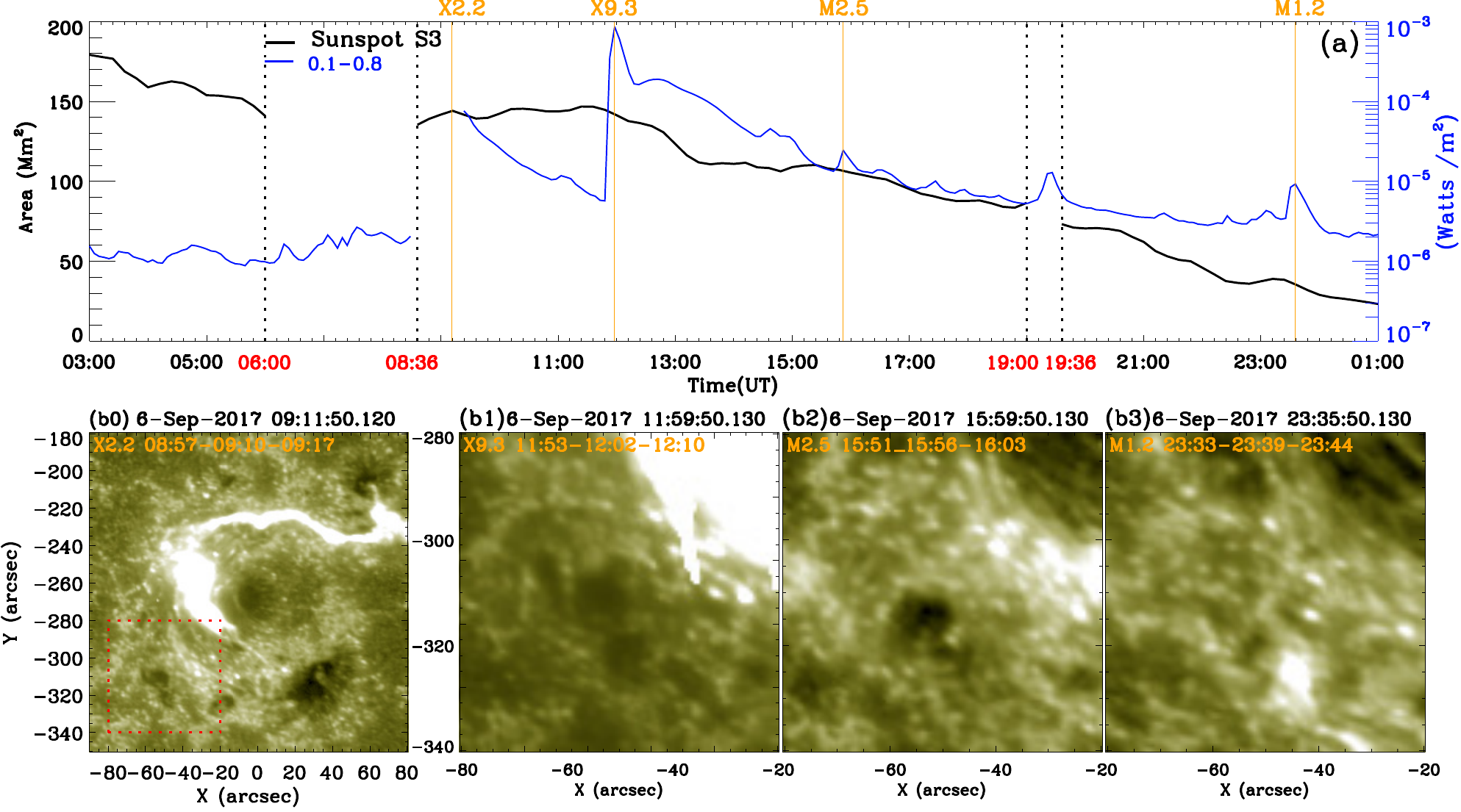}
\caption{(a):Variation of the total area in sunspot S3 from 03:00 UT on 2017 September 6 to 01:00 UT on September 7. The red line indicates the GOES X-ray flux profile of 0.1-0.8 nm in the corresponding time. (b0)-(b3): Temporal evolution of the sunspot S3 in the SDO/AIA 1600$\AA$ images. \textbf{The red dashed box in panel (b0) outlines the field of Region 3 (the view of panels (b1)-(b3)).}   \label{fig13b}}
\end{figure}

During the decay of the sunspot group, the area, the mean transverse magnetic field strength and the mean magnetic inclination angle in umbra and penumbra all decreased with time.  \textbf{The area in umbra kept a constant (around $23\, Mm^2$ ) from 03:00 UT to 12:00 UT on Sep.6. During this period, the mean transverse magnetic field strength in umbra remained at 600G and the mean magnetic inclination angle kept at $20\,^{\circ}$. After 12:00 UT, the area in umbra rapidly decreased from $20\, Mm^2$ at 12:00 UT on Sep.6 to around $5\, Mm^2$ at 01:00 UT on Sep.7. The mean transverse magnetic field strength in umbra decreased from 600G at 12:00 UT on Sep.6 to 400G at 01:00 UT on Sep.7. The mean magnetic inclination angle decreased from $20\,^{\circ}$ at 12:00 UT on Sep.6 to $10\,^{\circ}$ at 01:00 UT on Sep.7. For penumbra, the area decreased from $160\, Mm^2$ at 03:00 UT to $120\, Mm^2$ at 06:00 UT on Sep.6. Then it kept at $120\, Mm^2$ until 12:00 UT. Then the area in penumbra rapidly decreased from $120\, Mm^2$ at 12:00 UT on Sep.6 to around $20\, Mm^2$ at 01:00 UT on Sep.7. The mean transverse magnetic field strength in penumbra remained at 900G and the mean magnetic inclination angle kept at $32\,^{\circ}$ during 03:00 UT to 12:00 UT on Sep.6. After 12:00 UT, their value decreased (see Figure\ref{fig13}(a)-(c)).}

Figure \ref{fig13b} simply shows that the relationship between the evolution of S3 and the flare in active region. The blue curve in Figure\ref{fig13b}(a) indicates the GOES X-ray flux profile of 0.1-0.8nm during the evolution of S3. There were two X-flare in the NOAA 12673 during 08:36 UT to 12:10 UT. Although the big flare did not erupted in Region 3 (see the red box in Figure \ref{fig13b} (b0)), the rapid vary in the area and the transverse magnetic field of sunspots group occurred after the flare (after 12:00 UT as shown in Figure\ref{fig13}). The rapid decay of S3 may be related to the flare.

\begin{figure}
\plotone{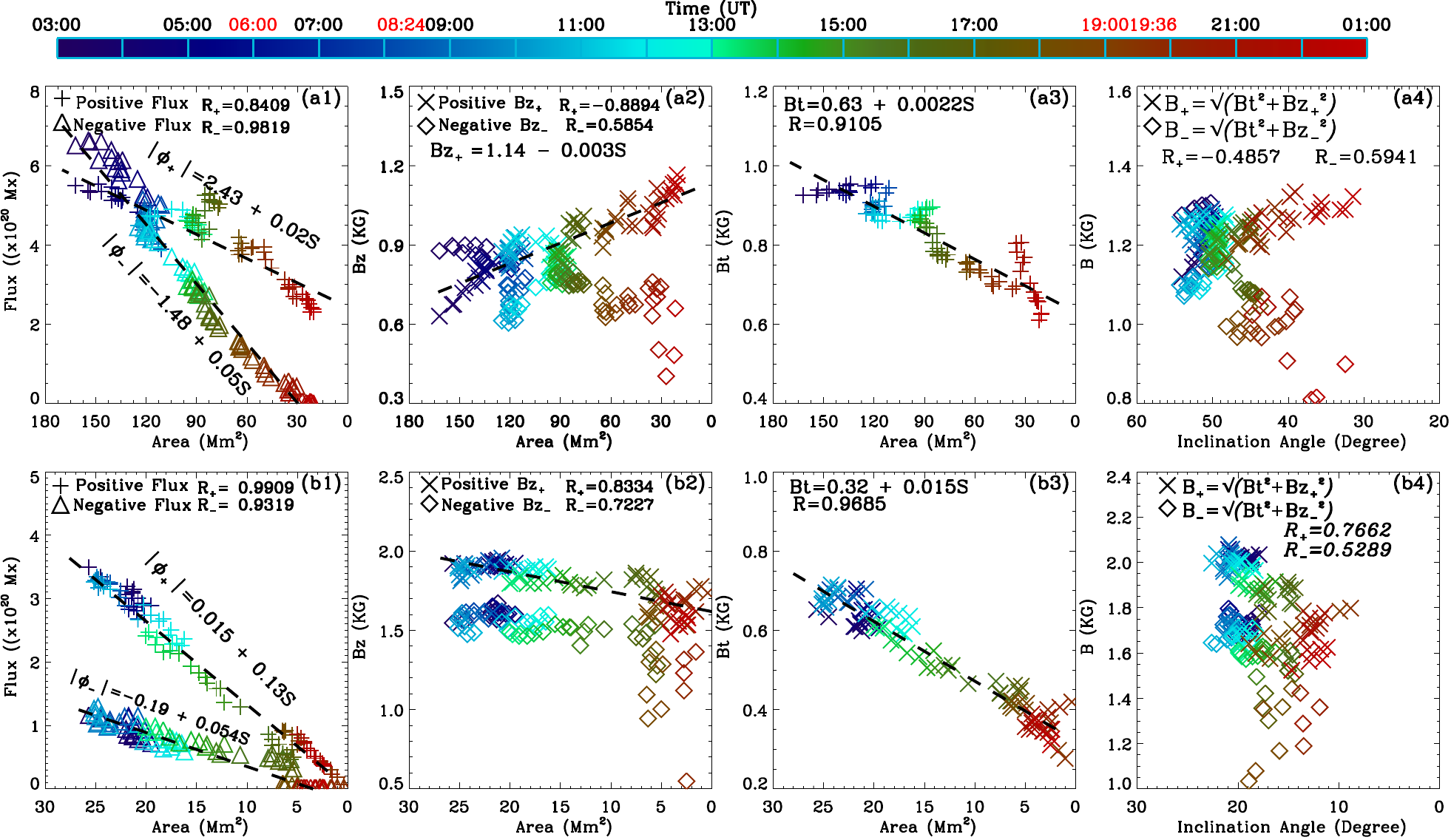}
\caption{Plots of the total magnetic flux, the mean longitudinal magnetic field strength (Bz) and the mean transverse magnetic field strength (Bt) for the penumbra/umbra of the sunspot S3 as a function of the penumbral area (top, (a1)-(a3))/the umbral area (bottom, (b1)-(b3)) of S3 during the decay of S3. Plot of magnetic field strength (B) for the penumbra/umbra of the sunspot S3 as a function of the mean magnetic inclination angle for penumbra (top, (a4))/ umbra (bottom, (b4)) of sunspot S3 during the decay of S3. The colours of symbols represent the evolution of time. \label{fig14}}
\end{figure}

Figure \ref{fig14} shows that the correlations of the area and the total magnetic flux/the mean longitudinal magnetic field strength (Bz)/the mean transverse magnetic field strength (Bt), the mean magnetic inclination angle and the mean total magnetic field strength in penumbra ((a1)-(a4)) and umbra ((b1)-(b4)) during the decay of the sunspot group.

Although the sunspots in the region 3 can not be distinguished in the continuum intensity images, the positive and negative magnetic flux of the sunspot group can be calculated separately. The ``$+$''symbols and ``$\bigtriangleup$'' symbols in Figure \ref{fig14}(a1)/(b1) respectively represent the positive and negative magnetic flux in penumbra/umbra of the sunspot group. The ``$\times$'' symbols and ``$\diamond$'' symbols in Figure \ref{fig14}(a2)/(b2) respectively represent the mean positive and negative longitudinal field strength in penumbra/umbra of the sunspot group.

During the decay of S3, the negative/positive magnetic flux in penumbra decreased with the decreasing penumbral area (see in Figure \ref{fig14}(a1)). In penumbra, the mean positive longitudinal magnetic field strength increased and the mean negative longitudinal magnetic field strength decreased with the decreasing penumbral area (see in Figure \ref{fig14}(a2)). The smallest value of the mean negative longitudinal magnetic field strength was 0 KG, we did not plot in the figure. This change tendency of the mean longitudinal magnetic field strength is consistent with the fact that the negative sunspot S3n almost completely disappeared and the positive one S3p became a small naked pore at the end of observational time. The mean transverse magnetic field strength decreased with the decreasing penumbral area (see in Figure \ref{fig14}(a3)). The magnetic field line in penumbra became more vertical during the decay of S3. The relationship between the mean total magnetic field strength and the magnetic inclination angle in penumbra are not notable, the correlation coefficient is small (Figure \ref{fig14}(a4)).

The negative/positive magnetic flux in umbra of S3 decreased with the decreasing umbral area (see in Figure \ref{fig14}(b1)). The mean positive/negative longitudinal magnetic field strength decreased with the decreasing umbral area. The correlation coefficient of the mean positive longitudinal magnetic field strength and the umbral area is higher than that of the mean negative longitudinal magnetic field strength and the umbral area (see in Figure \ref{fig14}(b2)). The mean transverse magnetic field strength in umbra decreased with the decreasing umbral area during the decay of S3 (see in Figure \ref{fig14}(b3)). The correlation between the mean total magnetic field strength and the magnetic inclination angle in umbra was no significant, the correlation coefficients are small (see in Figure \ref{fig14}(b4)). Basically, the mean total magnetic field strength and the magnetic field inclination in umbra gradually reduced during the decay of S3.

\begin{figure}
\plotone{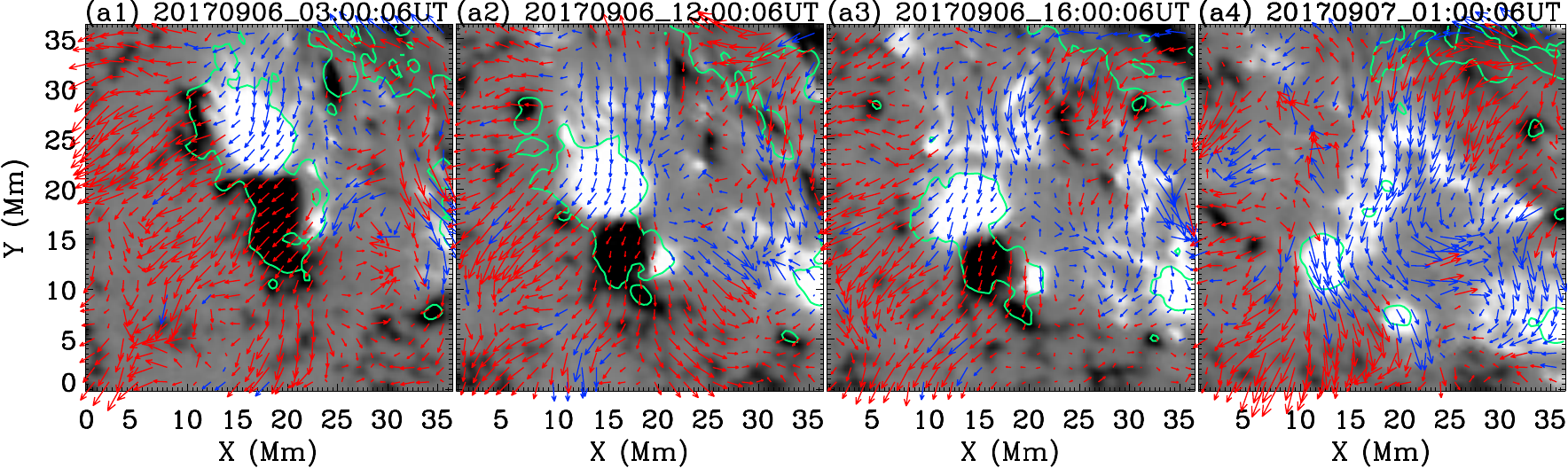}
\caption{ Horizontal velocity in the photosphere in Region 3 derived by DAVE4VM. The background image shows the vertical magnetic field with the positive field in white and negative field in black. The arrows represent horizontal velocity. Blue (red) arrows indicate that the vertical magnetic fields in the pixels are positive (negative). \label{fig14b}}
\end{figure}

Figure \ref{fig14b} (a1)-(a4)) show the variety of the horizontal motion around S3 during the decay of S3. The prominent outward flow around sunspot was not found in the surrounding of S3. In addition, the typical inflows in the inner and outward flow in the outer penumbra was absent in the region of S3 either. The strong emerging flow around S3 was dominant.

\section{Conclusion}\label{sec:conclusion}
\textbf{Three regions in the AR NOAA 12673 were chosen to analyze the formation and decay of sunspot penumbra and umbra. The main results are as follows:}

\textbf{1.The formation of penumbra in Sc and S2 was accompanied by continuous emerging flux. The penumbra of Sc and S2 first formed facing the side of flux emergence. The formation of penumbra was involved in two magnetic field systems: the continuous emerging magnetic field and the pre-existing magnetic field. The new emerging flux was trapped at the photosphere by the pre-existing magnetic field.}

\textbf{2.The magnetic field in the left penumbra of S1 gradually became more and more vertical during the disappearance of its left penumbra. The rearrangement in magnetic field caused the disappearance of left penumbra of S1 and the growth of right penumbra.}

\textbf{3.The penumbra of S2 on the opposite side of flux emergence disappeared first. The mean longitudinal magnetic strength in penumbra increased and the mean transverse magnetic strength in penumbra decreased with the decreasing penumbral area during the decay of S2. When sunspot S2 decayed, part of the horizontal magnetic lines in the penumbra became more vertical.}

\textbf{4.The decay of S3 was accompanied by magnetic cancellation. The abrupt vary in the area and the transverse magnetic field strength of S3 occurred after the X9.3 flare. }

\textbf{5.The dominant moat flow around sunspot S2 appeared gradually with the formation of penumbra. During the decay of penumbra, the outward flow first vanished on the side of magnetic emergence region and its speed gradually decreased.}

\section{Discussion}\label{sec:discussion}
Many previous results demonstrated that the formation of penumbra may be due to the emerging magnetic flux trapped at the photosphere. The pre-existing horizontal field in the chromosphere acts as a suppressor of the emerging flux \citep{Leka..1998ApJ...507..454L,Shimizu..2012ApJ...747L..18S,Romano..2013ApJ...771L...3R}. The importance of the overlying chromospheric field for the formation of the penumbra was demonstrated in observations \citep{Lim..2013ApJ...769L..18L} and the numerical simulations \citep{Rempel..2012ApJ...750...62R}. Our observations show that the penumbral formation of the sunspots may be formed by the interaction of two magnetic field systems. The penumbral filaments of sunspots formed from the later emerging magnetic flux that was trapped in the photosphere. The magnetic pressure from the already emerged magnetic field suppressed the later emerging flux.

\begin{figure}
\plotone{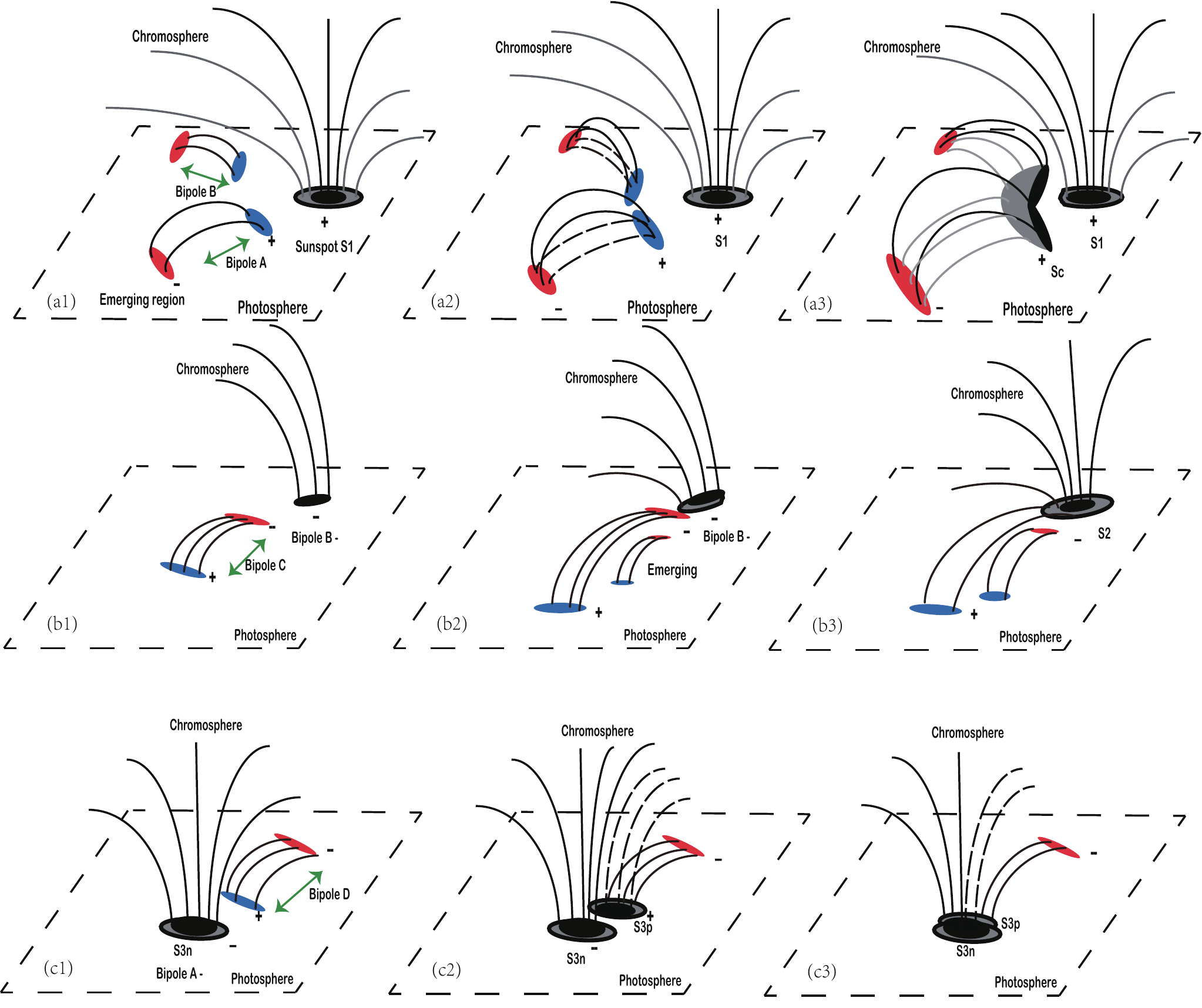}
\caption{Sketch of the magnetic field configuration during the evolution of sunspots in the three regions. (a1)-(a2): Region1, the disappearance of penumbra in S1 and the formation of penumbra in Sc. (b1)-(b3): Region 2, the formation of penumbra in  S2. (c1)-(c3): Region 3, the penumbral decay of sunspots S3n and S3p. \label{fig15}}
\end{figure}

\textbf{For our studied cases, the formation of penumbra in Sc and S2 accompanied by continuous emerging flux. The formation of Sc was involved in the continuous emerging magnetic field (Bipole A and Bipole B) and the pre-existing sunspot's magnetic field (S1). The magnetic pressure from the existence of magnetic field hindered the upward motion of the later emerging magnetic field. The ongoing emerging magnetic flux was trapped in the photosphere resulting in the penumbral formation of Sc. A cartoon is drawn to show the penumbral formation of Sc and the penumbral disappearance of S1. The blue and the red patches indicate the positive and negative polarities. The dark and the gray area indicate the umbra and penumbra of the sunspot (see in Figure \ref{fig15}(a1)-(a3)). The penumbral formation of sunspot S2 was also related to two magnetic field systems: the already emerged magnetic field systems (Bipole B) and the later emerging magnetic field (Bipole C) (see Figure \ref{fig15}(b1)). As the later emerging magnetic flux of Bipole C merged with the pre-existing small patches of Bipole B, the total magnetic flux of pore gradually increased and the outermost magnetic field lines of pore changed from vertical to horizontal. And then, the later emerged magnetic field was gradually trapped in the photosphere by the already emerged magnetic field systems.  A cartoon in Figure \ref{fig15}(b1)-(b3) show the formation of penumbra in S2.}

According to the results of \cite{Kitai..2014PASJ...66S..11K}, the sunspot penumbra can be formed by active accumulation of magnetic flux (the moving magnetic flux concentrations combined together to form a denser magnetic concentration), rapid emergence of magnetic fields (new magnetic flux of the same polarity), and twisted or rotating magnetic flux tubes (the penumbra are seen rotating with respect to the radial direction from the umbra center). The AR NOAA 12673 belongs to the way of rapid emergence of magnetic fields accumulation to form a sunspot: the rapid emergence of the small-scale bipole flux. The small-scale bipole that appear around the sunspot can give a significant contribute to the formation of sunspot penumbra. Due to the ongoing emerging flux of the bipole, the magnetic flux in the sunspot umbra and penumbra increased with the evolution time. As the two polarities of the bipole move away from each other, the magnetic lines connecting the bipole were elongated and become more and more horizontal. The penumbra can easily formed in the region between the bipole.

The role of the new emerging bipole flux in penumbral formation and decay may depend on its polarities and the polarity of the adjacent sunspot. There are three different situations: First, if the polarity of one of the bipole approaching the adjacent sunspot is same as that sunspot and they do not merge together, the penumbra filaments of the emerging polarity first appeared along the direction of its movement. The penumbra of the adjacent sunspot facing the emerging bipole will disappear (like the case in Region 1). Second, if the polarity of one of the bipole approaching the adjacent sunspot is same as that sunspot and they merge together, the penumbra filaments of the adjacent sunspot first appeared at the merging place (like the case in Region 2). Third, if the polarity of one of the bipole approaching the adjacent sunspot is opposite to that sunspot, it will cause the disappearance of penumbra between the emerging magnetic bipole and the adjacent sunspot (like the case in Region 3). Whatever the cases, the ongoing flux around the sunspot indeed affect the evolution of penumbra. For sunspot S2, the penumbra only partly encircled its umbra, the penumbra and moat flow in the side near the flux emergence was absent. It was hard for S2 to sustain its stability without a stable penumbra.

The rapid penumbral decay and umbral strengthening after solar flare are often observed \citep{Wang..2004ApJ...601L.195W,Deng..2005ApJ...623.1195D,Wang..2012ApJ...748...76W}. The impact of flare on the sunspot is rapid and could change the sunspot's area suddenly. The rapid rearrangement of sunspot magnetic field could induce the disappearance of sunspot's penumbra after solar flare.  \cite{Verma..2018A&A...614A...2V} focused on a decaying sunspot with the high-resolution observations. A darkened area with the same properties as umbra was found in the decaying penumbra.  They proposed that the horizontal magnetic fields in decaying penumbra became vertical. This process is same as the flare-induced rapid penumbral decay, only in a different time-scale. In our studies, the decay of penumbra is closely related to the rearrangement of its magnetic field lines.

\textbf{For the sunspot S1, the disappearance of penumbra in S1 stems from the change of magnetic field direction. When the newly emerging flux collided with S1, the leftmost magnetic field lines of S1 became more and more vertical from east to west. While the magnetic filed lines in the left sector of S1 became more vertical, it became more horizontal in the right sector of S1. The changes of magnetic field in sunspot's penumbra which induce by the flux emergence affect the development of sunspot. For the sunspot S2 and S3, the penumbral magnetic field also became more vertical during theirs decay. During the course of the penumbral disappearance, the mean magnetic inclination angle in penumbra of S2 decreased from $60\,^{\circ}$ to around $45\,^{\circ}$ and that of S3 decreased from $55\,^{\circ}$ to around $35\,^{\circ}$. A cartoon in Figure\ref{fig15}(c1)-(c3) show the disappearance of the penumbra of S3.}

\textbf{Moreover, the evolution tendency of area in S2 had no abrupt change after the series of flare eruptions. However, the flares could affect the magnetic field surrounding the sunspots and then it may affect the evolution of sunspots. Normally, a mature sunspot with umbra and penumbra lives for hours to months\citep{Hathaway...2008SoPh..250..269H}. The sunspot S2 has a shorter lifetime. As soon as the rudimentary structure of S2 was formed, S2 began to decay. The flares may affect the lifetime of sunspots. But it needs more studies to address this issue. After all, the ongoing flux emergence may also can affect the lifetime of S2. It is hard to differentiate the effect of the flux emergence and the surrounding magnetic field on the decay of S2. These unstable factors around sunspot likely changed the condition that are needed to sustain a stable sunspot.}

\cite{Leka..1998ApJ...507..454L} found a self-similarity between a growing pore and a small mature sunspot with respect to the magnetic field strength and the magnetic inclination angle distribution. In their study, the formation of the penumbra was preceded by an increase in the pore's magnetic field strength and the flux history of the pore-to-sunspot transition was not solely a function of pore's size, but initial intensification of the magnetic fields. \cite{Jur..2011A&A...531A.118J} discussed the magnetic inclination angle and the magnetic field strength, the sunspot area and the magnetic field strength, and the sunspot area and the magnetic inclination angle distributions on penumbral boundaries of the mature sunspots. The magnetic field strength and inclination on the outer penumbral boundary was decreasing with the decreasing sunspot area. The vertical component of magnetic field is independent of the sunspot area. Along the inner penumbral boundary, both of the magnetic field strength and inclination on average decreased with the decreasing umbral area. Compared to large umbrae, the boundaries of small umbrae have weaker and more vertical magnetic field.

In our study, we presented the area and the magnetic flux, the area and the longitudinal magnetic field strength, the area and the transverse magnetic field strength, and the magnetic inclination angle and the magnetic field strength distributions in penumbra and umbra during the evolution of three sunspots. During the formation of sunspot, both the mean magnetic flux and the mean transverse magnetic field strength in penumbra and umbra increased with the increasing area. The increase of penumbral area accompanied by a gradual decreasing longitudinal magnetic field. Instead the growth of umbra accompanied by an increasing longitudinal magnetic field. The total magnetic field strength in penumbra almost kept a constant with the increasing mean magnetic inclination angle. The total magnetic field strength in umbra increased with the increasing mean magnetic inclination angle. For the case of decaying sunspots, the mean magnetic flux and the mean transverse magnetic field strength in penumbra and umbra decreased with the decreasing area. The decay of the penumbra accompanied by an increasing longitudinal magnetic field. The decay of the umbra occurred with a gradual decreasing longitudinal magnetic field.

Whether the penumbra developed at the expense of umbral magnetic flux remains contradictory. \cite{Jur..2015A&A...580L...1J} proposed that the penumbra developed at the cost of the pore magnetic flux, which supports the result suggested earlier by \cite{Watanabe..2014ApJ...796...77W}. They found the umbral area of sunspot decreased during the formation of its penumbra. However, \cite{Schlichenmaier..2010A&A...512L...1S} found that the umbral area remained a constant value during the formation of a sunspot penumbra. In our study, both the umbral area and penumbral area of sunspots increased during the penumbral formation of the sunspot. The development of penumbra was not at the expense of umbral magnetic flux due to that the continuous flux emergence can provide sufficient magnetic flux for the development of penumbra. However, the magnetic flux in umbra of S2 decreased while the magnetic flux in penumbra increased during the decay of S2. Simultaneously, the umbral area decreased but the penumbral area developed into its maximum. The penumbra may develop at expense of the umbral magnetic flux. After the umbra completely separated into two part by the light bridge, the area and the magnetic flux of umbra recovered a little when the area and the magnetic flux in penumbra decreased. The recovery of umbra may be costing the penumbral magnetic flux. The transformation of magnetic flux between umbra and penumbra is found at the beginning of the sunspot decay.

\cite{Murabito..2017ApJ...834...76M} suggested that the rising velocity of magnetic flux bundle may affect sunspot structure and evolution. The complexity of emerging magnetic field in pore's surrounding may also influence the development of penumbra. \cite{Botha..2011ApJ...731..108B} presented a study of the decay process of large magnetic flux tubes and found that the decay of central flux tube rely on its surrounding convection. The convection around large sunspots can change the shape of the central flux tube. The convection, the rising velocity and complexity of the emerging magnetic flux may influence the formation and decay of sunspot. However, how does the emerging magnetic flux or convection affect the penumbral evolution? The answer is still open. We need more observations to confirm the correlation between the penumbra formation and the surrounding flux emergence. In the future, we will analyze more examples to address this issue by using the high-resolution observations from the NVST and other instruments.

We would like to thank the NVST, SDO/AIA, and SDO/HMI teams for the high-cadence data support. This work is sponsored by the National Science Foundation of China (NSFC) under the grant numbers 11873087, 11603071, 11503080, 11633008 by the Youth Innovation Promotion Association CAS (No.2011056), by the Yunnan Science Foundation of China under number 2018FA001, by Project Sup-ported by the Specialized Research Fund for State Key Laboratories and by the grant associated with project of the Group for Innovation of Yunnan Province.

\end{document}